\documentclass[%
 reprint,
superscriptaddress,
 amsmath,amssymb,
 aps,
prb,
]{revtex4-1}

\usepackage{graphicx}% Include figure files
\usepackage{dcolumn}% Align table columns on decimal point
\usepackage{bm}% bold math
\usepackage{upgreek}
\usepackage{notoccite}
\begin{document}

%\preprint{APS/123-QED}

\title{Asymmetric Magnetic Relaxation behavior of Domains and Domain Walls Observed Through the FeRh First-Order Metamagnetic Phase Transition}% Force line breaks with \\
\author{Jamie R.~Massey}
\email{jamie.massey@psi.ch}
\affiliation{School of Physics and Astronomy, University of Leeds, Leeds LS2 9JT, United Kingdom.}

\author{Rowan C.~Temple}
\affiliation{School of Physics and Astronomy, University of Leeds, Leeds LS2 9JT, United Kingdom.}

\author{Trevor P.~Almeida}
\affiliation{SUPA, School of Physics and Astronomy, University of Glasgow, Glasgow, G12 8QQ, United Kingdom.}

\author{Ray Lamb}
\affiliation{School of Physics and Astronomy, University of Glasgow, Glasgow, G12 8QQ, United Kingdom.}

\author{Nicolas A.~Peters}
\affiliation{School of Physics and Astronomy, University of Leeds, Leeds LS2 9JT, United Kingdom.}
\affiliation{School of Electronic and Electrical Engineering, University of Leeds, Leeds, LS2 9JT, United Kingdom.}

\author{Richard P.~Campion}
\affiliation{School of Physics and Astronomy, University of Nottingham, Nottingham, NG7 2RD, United Kingdom.}

\author{Raymond Fan}
\affiliation{Diamond Light Source, Chilton, Didcot, OX11 0DE, United Kingdom.}

\author{Damien McGrouther}
\affiliation{SUPA, School of Physics and Astronomy, University of Glasgow, Glasgow, G12 8QQ, United Kingdom.}

\author{Stephen McVitie}
\affiliation{SUPA, School of Physics and Astronomy, University of Glasgow, Glasgow, G12 8QQ, United Kingdom.}

\author{Paul Steadman}
\affiliation{Diamond Light Source, Chilton, Didcot, OX11 0DE, United Kingdom.}

\author{Christopher H.~Marrows}
\email{c.h.marrows@leeds.ac.uk}
\affiliation{School of Physics and Astronomy, University of Leeds, Leeds LS2 9JT, United Kingdom.}

\date{\today}% It is always \today, today,

\begin{abstract}
The phase coexistence present through a first-order phase transition means there will be finite regions between the two phases where the structure of the system will vary from one phase to the other, known as a phase boundary wall. This region is said to play an important but unknown role in the dynamics of the first-order phase transitions. Here, by using both x-ray photon correlation spectroscopy and magnetometry techniques to measure the temporal isothermal development at various points through the thermally activated first-order metamagnetic phase transition present in the near-equiatomic FeRh alloy, we are able to isolate the dynamic behavior of the domain walls in this system. These investigations reveal that relaxation behavior of the domain walls changes when phase coexistence is introduced into the system and that the domain wall dynamics is different to the macroscale behavior. We attribute this to the effect of the exchange coupling between regions of either magnetic phase changing the dynamic properties of domain walls relative to bulk regions of either phase. We also believe this behavior comes from the influence of the phase boundary wall on other magnetic objects in the system.
\end{abstract}
\maketitle

%\tableofcontents

\section{Introduction}
The dynamic behavior of second-order phase transitions is well understood, owing largely to the critical scaling of the correlation length of thermal fluctuations approaching the temperature associated with the phase transition \cite{Porter, Djurberg1997, Morley2017}. However, due to the presence of latent heat, the same behavior is not expected through first-order phase transitions \cite{Porter}, meaning that the study of first-order phase transition dynamics remain a topic of active investigation \cite{McLeod2019, Nemoto2017, Post2018, Keavney2018, Kundu2020}. Recent breakthroughs in imaging techniques, capable of tracking various material properties, has led to a surge of interest in materials that exhibit first-order phase transitions \cite{McLeod2019, Post2018, Keavney2018}. These investigations focus on the quasi-static development of the first-order phase transition dynamics. They show that first-order phase transition systems demonstrate critical scaling behavior of both the domain size \cite{Keavney2018, McLeod2019}, and the phase boundary wall \cite{Post2018, McLeod2019}, which acts to blur the once definitive line between the two phase transition classifications. The coupling between regions of the two phases is cited as the source of this quasi-second-order behavior \cite{Nemoto2017, Keavney2018, McLeod2019, Post2018, Kundu2020}. 

Another interesting aspect of phase transition dynamics is their temporal relaxation behavior \cite{Morley2017,Chen2018, Kundu2020}. Recently, quasi-second-order behavior has been observed in the phase-ordering, and relaxation times in a Mott insulator-metal transition system \cite{Kundu2020}. Despite this wave of recent interest in first-order phase transition dynamics, the role of the phase boundary wall in these proceedings remains unclear. Aside from the critical scaling of the size of the phase boundary wall approaching the transition temperature \cite{McLeod2019, Post2018}, very little is known about this region. It is said to play a key, but as yet unknown, role in the evolution of the first-order phase transitions \cite{Maat2005, Moriyama2015}.

X-Ray Photon Correlation Spectroscopy (XPCS) has been used to study the relaxation dynamics of domain walls in magnetic systems \cite{Sinha2014,Morley2017,Chen2018,Shpyrko2007, Konings2011, ChenJam2013}. The FeRh alloy is a material that, in a specific composition range \cite{Swart1984}, undergoes a coupled first-order magnetostructural phase transition from an antiferromagnetic (AF) to a ferromagnetic (FM) state when heated through a transition temperature, $T_\text{T}$, that is $\sim 380$~K in bulk \cite{Fallot1939, Kouvel1962, Kouvel1966, devries2013}. The coupled magnetic, charge and structural transitions \cite{Ricodeau1972, Thiele2004, Ju2004, Radu2010} in this material make it an ideal candidate for use in a plethora of possible magnetic data storage architectures \cite{Thiele2003,Cherifi2014,Lee2015,Marti2014,LeGraet2015,Moriyama2015,Temple2019}. Here, by comparing measurements of the isothermal relaxation behavior through the phase transition performed using XPCS and magnetometry techniques, we are able to isolate the relaxation behavior of the domain walls. These investigations reveal that the dynamic behavior of domain walls where phase coexistence is present is different to both the FM/AF domain walls and the nucleation and growth of magnetic domains. We believe this behavior emanates from the influence of the phase boundary wall on other objects in the system and that the change in behavior compared to other objects is due to the influence of interphase exchange coupling that accompanies the phase coexistence in this system. 

\section{Experimental Details}
\subsection{Sample Growth and Characterization}
\begin{figure}[t]
\includegraphics[width = 7cm]{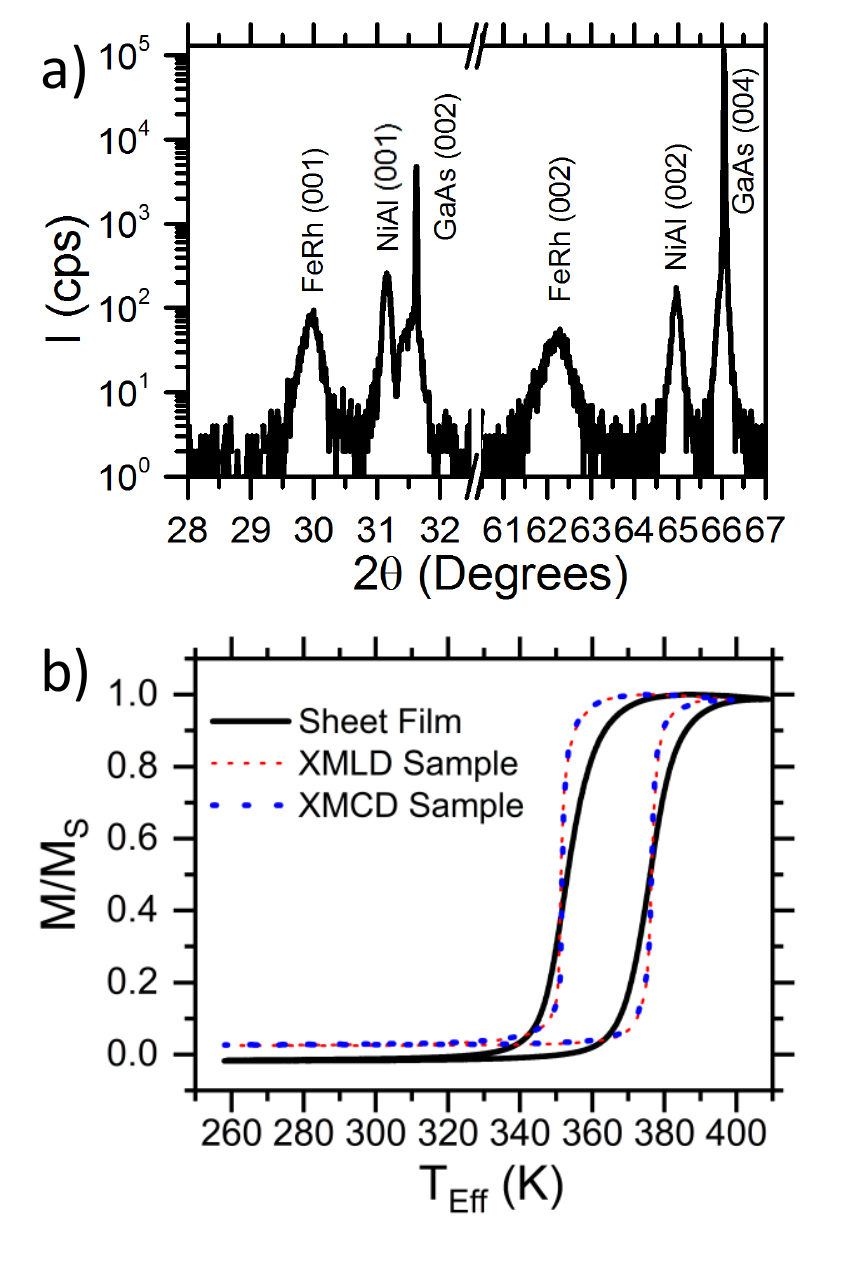}
\caption{FeRh thin film sample characterization. (a) Ambient temperature XRD scan with indexed Bragg peaks. The presence of both the (001) and (002) reflections for both NiAl and FeRh indicates the presence of chemical order in both layers. (b) Magnetometry traces plotted against $T_\text{Eff}$ through the range of the transition taken with a 1~T field applied in the film plane. The black line shows the behavior of the as-grown sample whilst still attached to the substrate, whilst the red and blue lines show the sample behavior after being made into a membrane. These two lines show samples that were used in the XMLD and XMCD experiments, respectively, which are close to indistinguishable.
\label{fig:x1}}
\end{figure}

The sample used in this experiment was a 100~nm thick FeRh layer grown on a 100~nm thick NiAl buffer layer deposited using DC magnetron sputtering on a molecular beam epitaxy-grown GaAs(25 nm)/AlAs(25 nm)/GaAs heterostructure substrate. NiAl is also a B2-ordered material and the layer used here improves the stability of the FeRh layer and promotes epitaxial growth \cite{Kande2011}. The substrate was annealed overnight at 400$^\circ$C, at which temperature the NiAl layer was deposited. The system was then heated to 600$^\circ$C, where the FeRh layer was grown. The sample was then annealed \textit{in situ} for 1 hour at 700$^\circ$C. Structural characterization of the as-grown film was performed using ambient temperature x-ray diffraction (XRD) and is shown in Fig.~\ref{fig:x1}(a). The observation of the (001) and (002) reflections for both the NiAl and FeRh layers demonstrates chemical order in both layers. Further analysis yields values of the room temperature lattice constant, which have been averaged across both of the peaks present in Fig.~\ref{fig:x1}(a), $a = 2.981 \pm 0.004$ \AA~ and $a = 2.869 \pm 0.001$ \AA~ for FeRh and NiAl respectively. 

The expected magnetic transition of the as grown sample is evident when measured using a SQUID-vibrating sample magnetometer (SQUID-VSM) in a 1~T in-plane applied field, shown by the black line in Fig.~\ref{fig:x1}(b). The transition temperature of FeRh is sensitive to the application of external field \cite{Maat2005}. Therefore, to be able to directly compare the magnetometry with the XPCS measurements (shown in the Results section) which are performed without an external field applied, this figure is plotted against the effective temperature, $T_\text{Eff}$. The latter is the temperature at which the same magnetization is expected to be achieved in the absence of external field (see supplementary material for calculation) \cite{Maat2005, Massey2019}.

As the magnetic domains in FeRh are of $\mu$m dimensions \cite{ Almeida2017, Temple2018}, the $Q$ range in which they are found falls within the regime only accessible through transmission experiments. As such, the as-grown film was made into a membrane suitable for x-ray transmission measurements using a HF etching process. This was performed following the method outlined in Ref.~\onlinecite{Russell2013}. The substrate was chosen for its known etching chemistry. By performing the etch in this manner, it is possible to destroy the AlAs layer without harming the rest of the sample. This creates a free standing FeRh(100 nm)/NiAl(100 nm)/GaAs(25 nm) layer, which was subsequently captured between two Cu TEM grids to provide an x-ray transparent sample of B2-ordered FeRh.

After undergoing the etching process to be made into a soft x-ray transparent membrane the phase transition of the membrane sample was measured in a 1~T magnetic field applied within the film plane, shown by the coloured lines in Fig.~\ref{fig:x1}(b). There are two membrane samples here, one used for the experiments concerned with each type of x-ray magnetic dichroism, namely X-ray Magnetic Circular Dichroism (XMCD) and X-ray Magnetic Linear Dichroism (XMLD), which will be explained in more detail in the results section. The two come from the same parent film it is clear from Fig.~\ref{fig:x1}(b) that the magnetic transition is still present in the membrane samples and has become considerably sharper. As the magnetic transition of the as-grown and membrane samples were measured using the same conditions, the difference in the behavior of the two samples comes as a result of removing the substrate, and therefore the strain on the film from the lattice mismatch with the substrate. The two membrane samples have indistinguishable metamagnetic transitions.

\subsection{Soft X-Ray Methods}
The XPCS measurements were carried out at the I10 beamline of the Diamond Light Source. A 20~$\upmu$m radius pinhole is placed in the beampath in front of the samples to provide the coherent light required to generate the speckle pattern \cite{Sinha2014,Morley2017}. To increase the scattering from the magnetic parts of the sample, these measurements were performed with photon energies at the Fe $L_3$ resonance. This is measured to be between 706.4-707.2~eV at 400~K depending on the type of dichroism used. The measurements shown in this work were collected over 3 beamtimes, two of which focus on the XMCD measurements and the third which focusses on XMLD. The beam energy used in the three beamtimes were 707, 707.4 and 706.2 eV, respectively. The characterization of the Fe L$_3$ edge for this sample, and the position of the energy used in each beamtime relative to this, can be seen in the supplementary information. 

The image series were taken using a 2D charge coupled device (CCD) camera. For each XPCS image series, in which consecutive images are performed with opposing helicities (XMCD) or polarization orientations (XMLD) and are combined in post-processing to increase the signal as per the method outlined by Fischer \textit{et al}. \cite{Fischer1997}. In this protocol, the intensity of each image of the final image, $I$, is calculated using
\begin{equation}
I = I^+ - I^-,
\end{equation}
where $I^+$ is the intensity of the image taken with a given helicity or linear polarization and $I^-$ is the image taken using the opposite helicity or linear polarization. An example of one of these images, taken after cooling to 390 K using circularly polarized light, can be seen in Fig.~\ref{fig:SAXS}(a). 

Each image series consists of 100 images calculated in this way and are taken over 90-120 minute periods, depending on the type of dichroism used. The sample was thermally cycled between each measurement to reset the domain structure. For measurements performed on the cooling arm, the system was cycled into the fully FM phase (400 K) and then cooled to the desired temperature. For measurements on the heating branch, the system was cycled into the fully AF phase (270 K) and then heated to the desired measurement temperature. All XPCS measurements took place in the absence of a magnetic field. Also included in the data are measurements performed using only a single helicity of circularly polarized light. This occurred after an issue with the undulator during the first beamtime. The signal is much weaker for these measurements than it is for the measurements where the XMCD protocol can be used to boost the signal. These measurements use between 40-90 images taken approximately 1 minute apart and are included in this analysis as the values extracted from the fitting of the $g_2$ equation are similar to the measurements with larger signals. The value of $A$ extracted for the measurements in which a single helicity is used is not included, as it is significantly lower than the two helicity measurements.

\section{Results}
\subsection{Structural Analysis}
\begin{figure}[t]
    \includegraphics[width = 8cm]{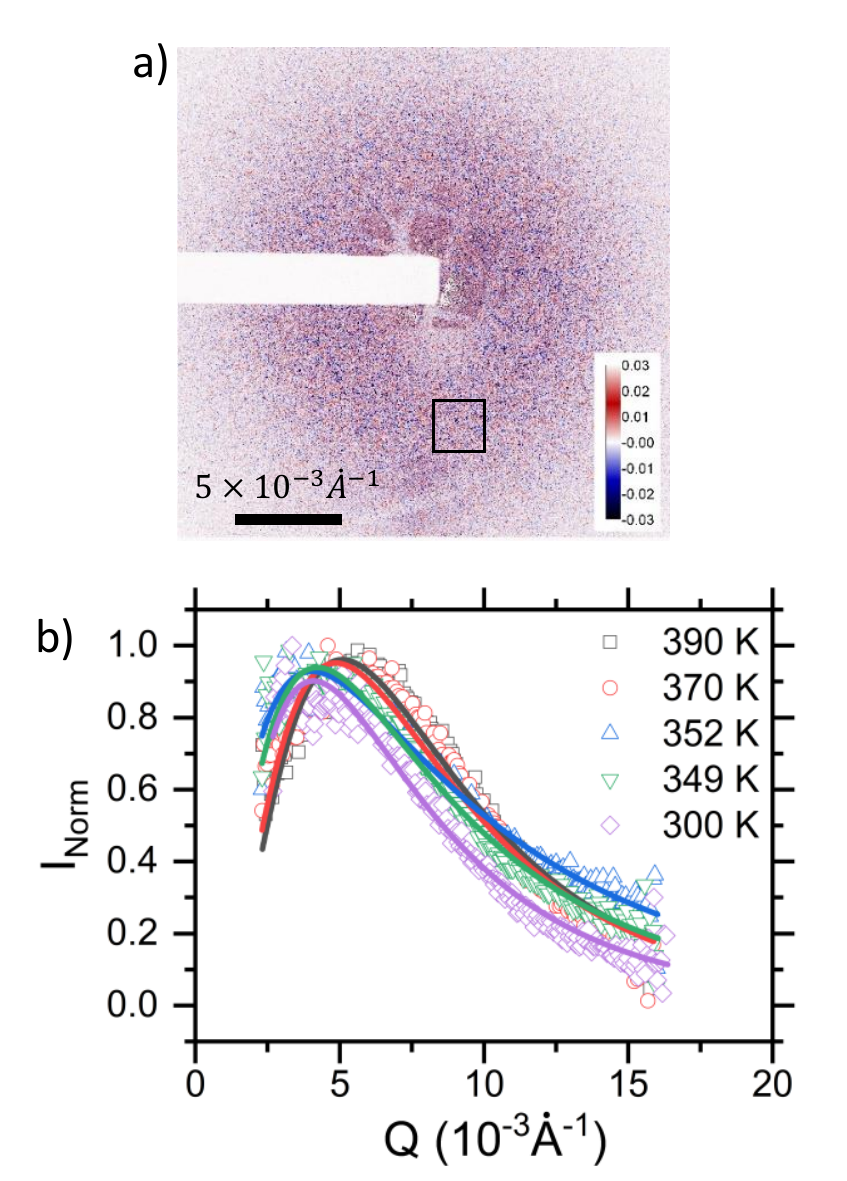}
    \caption{Resonant magnetic X-Ray scattering. (a) Image taken on the Fe L$_3$ resonance edge using circularly polarized light after cooling to 390 K, showing a clear small angle scattering ring with speckle. The shadow cast by the beamstop and some direct transmission though the pinhole and crossed TEM grids is visible in the centre of the image. The colour here represents the intensity of the difference image, with blue regions having a higher intensity of one polarization, whilst the red regions have a higher intensity from the other. (b) Examples of the radial intensity profiles (points) and the fit of a log-normal distribution to the data (lines) taken at various temperatures on the cooling branch using linearly polarized light.}
\label{fig:SAXS}
\end{figure}

To track the behavior of the magnetic structure within the scattering ring the first image of each series was taken and its radial average was calculated. This procedure was performed after any potential artefacts within the image, such as the grid, the back of the camera and any holes in the beamstop are removed. $Q = 0$ is taken to be the centre of mass of the image. The XPCS measurements of the dynamic behavior were performed on a $200 \times 200$ pixel box centered around the peak in structural analysis, an example of which can be seen in Fig.~\ref{fig:SAXS}(a).

\begin{figure}[t]
    \includegraphics[width = 8cm]{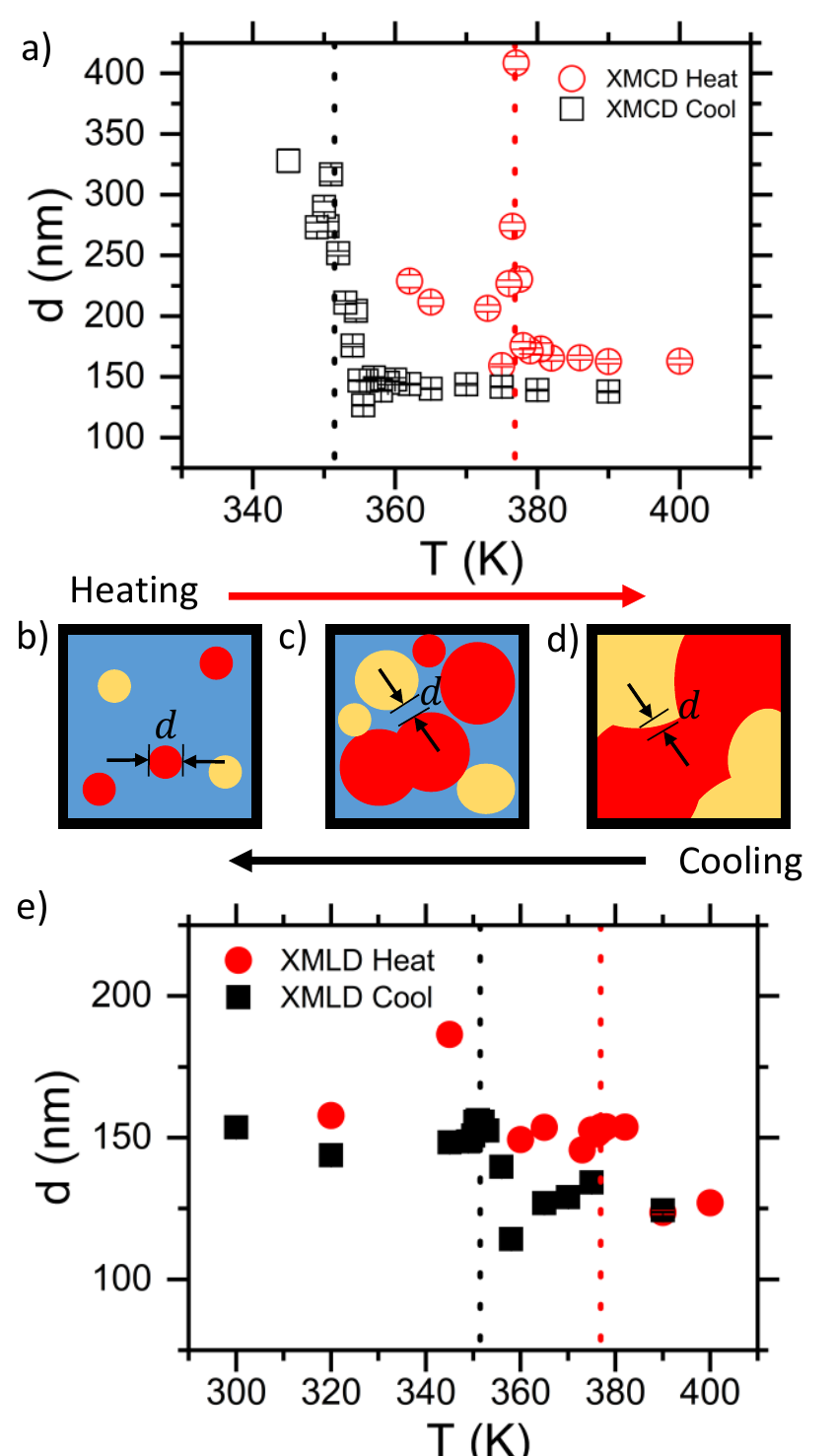}
    \caption{Temperature dependence of the peak in the radial distribution. (a) and (e) show the length $d$ that corresponds to the position of the peaks seen in the radial intensity profiles, $Q_\text{Peak}$, against $T$ for measurements performed using circularly (labelled XMCD) and linearly (labelled XMLD) polarized light respectively. The dashed lines in these panels show the position of the transition midpoint, $T_\text{M}$, for each branch, extracted from the magnetometry. Note the different $T$ scales on the abscissae of the two graphs. Panels (b) - (d) show cartoons of the magnetic state at various points through the transition. In these diagrams, the blue depicts AF material and the red and yellow regions are FM domains with the magnetization oriented in opposite directions. The arrows are used to indicate the temperature sweep direction.}
\label{fig:SAXS_results}
\end{figure}

Examples of these radial average of the intensity profiles $I(Q)$ for the first image of each of the XPCS sets, where $Q$ is the wavevector transfer, can be seen for various temperatures when cooling performed using linearly polarised light in Fig.~\ref{fig:SAXS}(b). This reveals a peak which corresponds to correlations in the structure factor \cite{Sivia, Bagschik2016}. These are presented in normalised form as $I_\text{Norm}(Q) = (I(Q) - I_\text{Min})/(I_\text{Max} - I_\text{Min})$. A log-normal distribution was fitted to each data set to identify the position of the peak \cite{Fischer1997}, $Q_\text{Peak} = Q_0e^{-\omega^2}$, where $Q_0$ is the centre of the distribution and $\omega$ is the logarithm of the full width at half maximum (FHWM) of the peak. It is then possible to extract the spatial lengthscale associated with the peak, $d = 2\pi/ Q_\text{Peak}$, the results of these fittings for all measurements taken using both circular and linear polarisation are shown in Fig.~\ref{fig:SAXS_results}. By using both both XMCD, which is sensitive to the behavior of the FM domains only, and XMLD, which is sensitive to the orientation of the spin-axis of the material and can access both FM and AF order \cite{stohrbook}, we are able to access the behavior of both the AF and FM phases, giving a more holistic understanding of the system.

Fig.~\ref{fig:SAXS_results}(a) shows measurements taken using circular light for both heating and cooling branches. These reveal an increase towards a peak value at $d \sim 400$~nm when heating, which occurs at around 377~K. The dashed lines here mark the position of the transition midpoint, $T_\text{M}$, calculated as the steepest point in the magnetometry trace in Fig.~\ref{fig:x1}(b). This convention is used throughout this work and should be taken to be the case unless specified otherwise. At temperatures in excess of $T_\text{M}$, $d$ is seen to decrease again in Fig.~\ref{fig:SAXS_results}(a), falling to 150~nm at 400~K. A similar behavior is seen when cooling, though $d$ is seen to be constant at around 150~nm until the temperature falls below 350~K, after which it is seen to increase, also to about 300~nm, before just starting to drop. These measurements performed during cooling also see the peak value occurring for $T = T_\text{M}$. In order to understand these findings it is necessary to consider the development of the magnetic domain structure through the transition.

Previous real-space imaging has shown that FM domains in FeRh nucleate as flux closed structures around 200~nm in diameter on heating out of the fully AF phase \cite{Almeida2017,Temple2018}. As the transition progresses, these domains begin to agglomerate, surrounding small residual patches of AF material before the film becomes fully FM with the domain structure taking up a striped configuration in the absence of a magnetic field, where the stripes have $\mu$m dimensions \cite{Almeida2017, Temple2018}. Scattering from objects of this size would fall outside the available $Q$ range, however scattering across the domain wall between the domains may be accessible. In this system, there are two types of domain wall scattering to consider: i) that between adjacent FM domains and ii) those between domains that are separated by regions of AF material. The lengthscale associated with scattering from these two objects would have different temperature dependences through the transition. Given the limited $Q$ range and the changing nature of the domain structure in this experiment, it may be expected that the nature of the scatterer changes across the transition.

To help aid the discussion, diagrams of the magnetic states through the transition are included in panels (b)-(d) in Fig.~\ref{fig:SAXS_results}. The blue regions depict regions of AF material, whilst the red and yellow regions are FM domains with their magnetization aligned in opposite directions. The black bars are used to indicate the source of the scattering object at each stage. 

Considering the measurements performed using circular light in Fig.~\ref{fig:SAXS_results}(a), when $T < T_\text{M}$, the measured size of the scattering object $d$ is consistent with the size of the FM domains seen in these previous works \cite{Baldasseroni2012,Almeida2017,Temple2018}. It is therefore reasonable to think that the scattering objects are the FM domains that have nucleated from the AF phase as the transition begins, which is shown by Fig.~\ref{fig:SAXS_results}(b). On heating, it is known that the size of the domains would increase, which is reasonable up to $T_\text{M}$ but is contradictory to the behavior seen here at higher temperatures. Nevertheless, is also known that there will still be regions of AF material present between these FM domains \cite{Baldasseroni2012,Kinane2014,Mariager2013,Almeida2017,Temple2018}, which will shrink when approaching the fully FM phase, decreasing the distance between FM domains. These gaps between the FM phase regions now become the scattering objects, as shown in Fig.~\ref{fig:SAXS_results}(c). When cooling from high temperatures, $d$ is invariant down to about 360~K. These temperatures are consistent with the fully FM phase when measured using magnetometry and so the scattering in this regime is believed to correspond to domain walls between FM regions with different magnetization directions, as seen in Fig.~\ref{fig:SAXS_results}(d). Approaching the transition, the rise and fall in $d$ as it passed through a peak at $T_\text{M}$ corresponds to the processes on the heating branch in reverse.

Turning to the measurements using linear light in Fig.~\ref{fig:SAXS_results}(e), the overall behavior is consistent, with peaks in $d$ appearing at $T_\text{M}$ on each branch, although it is less pronounced on the cooling branch. The length scales extracted for $T > T_\text{M}$ appear to be consistent with the corresponding points measured using circular light. Therefore, the nature of the scatterers in this region is believed to be the same in both measurements. However, the behavior of the two diverges for $T< T_\text{M}$, where the measurements taken using linearly polarized light appear to be constant at $d \approx 150$ nm down to temperatures of 300~K for both transition branches. This behavior is consistent with that seen in the only other previous magnetic imaging experiments on the AF phase of FeRh \cite{Baldasseroni2015} and hence this change in $d$ is attributed to a change in scatterer at $T_\text{M}$. At 300 K, the system would exhibit only a 4 \% FM volume fraction. Therefore as XMLD is sensitive to the presence of AF materials, the change in scatterer at this stage is believed to be AF domain structures, mirroring the FM domain structures causing scattering at the highest temperatures for the circular light measurements. 

\subsection{X-Ray Photon Correlation Spectroscopy}
\begin{figure}[t]
\includegraphics[width = 7cm]{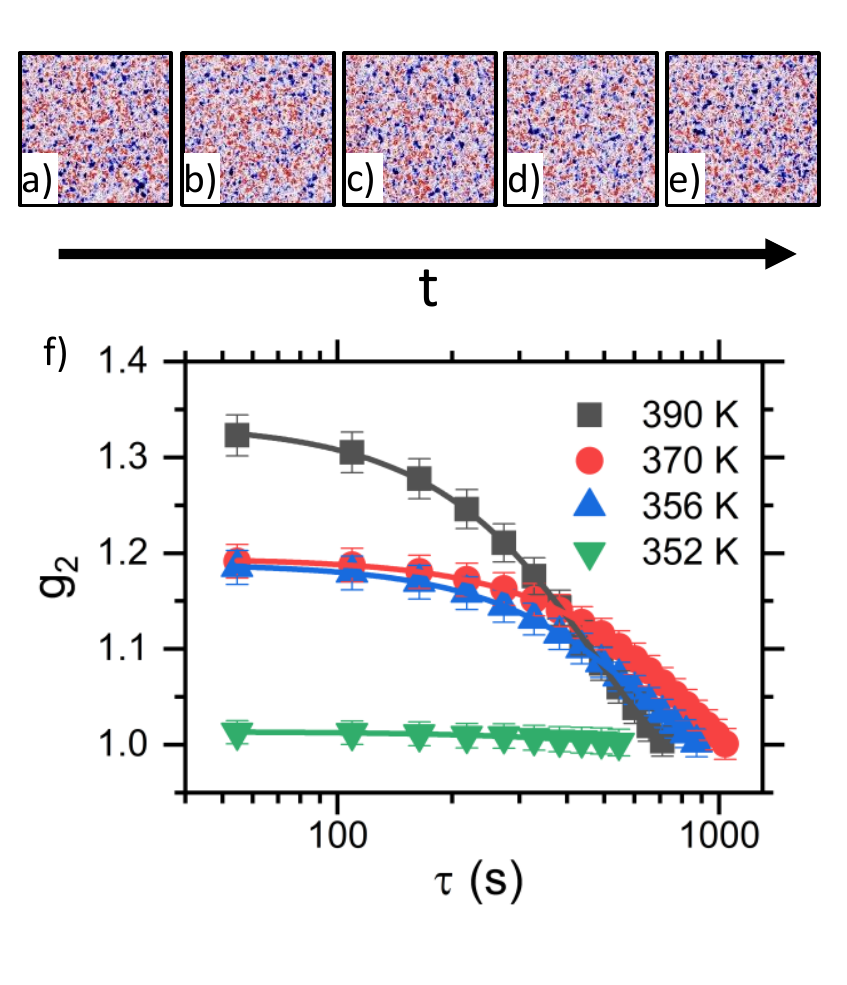}
\caption{Temporal correlation analysis.  (a) - (e) Example $200 \times 200$ pixel images where the temporal correlation analysis was performed taken at $t = 0$, 1310.4, 2620.8, 3931.2 and 5241.6~s for the measurement performed when cooling to 390 K using XMCD. The first image of this series is also shown in Fig.~\ref{fig:SAXS}(a). The speckle pattern clearly changes through the measurement time indicating the presence of dynamic behavior. The black square in panel (a) marks the approximate position where these images were taken. (f) Example $g_2$ functions derived from the images (points) and the fits to equation \ref{eq:g2fit} (lines) for various points on the cooling branch of the phase transition taken using circularly polarized light. \label{fig:x2}}
\end{figure}

XPCS follows the temporal correlations of fine structure present in diffraction features, known as speckle \cite{Sinha2014}. Panels (a)-(e) of Fig.~\ref{fig:x2} shows example speckle patterns taken through the measurement time within the black box shown in Fig.~\ref{fig:SAXS}(a). The changing speckle pattern is indicative of a varying magnetic state through the measurement time. Variations in the speckle intensity are due to fluctuations in the domain walls in magnetic systems \cite{ChenJam2013, Morley2017, Chen2018}.

To quantify the extent of these changes the temporal auto-correlations for each image series performed at a given temperature are calculated using a $g_2$ function. This function is calculated for a series of images separated by a time delay $\tau$ and takes the form \cite{Shpyrko2007,Sinha2014,ChenJam2013,Morley2017,Chen2018}
\begin{equation} \label{eq:g2}
g_2 (\tau) = \bigg \langle \frac{\langle I(\mathbf{Q}, t) I(\mathbf{Q}, t+\tau)\rangle_t}{\langle I(\mathbf{Q}, t)\rangle^2_t} \bigg \rangle_{\mathbf{Q}}
\end{equation}
where $I(\mathbf{Q}, t)$ is the intensity at position $\mathbf{Q}$ at time $t$ and where $\langle ... \rangle_{t, \mathbf{Q}}$ denotes an average over $t$ or $\mathbf{Q}$. Here, the $g_2$ function is calculated for each individual pixel, which corresponds to a particular value of $\mathbf{Q}$, each of which is then averaged over the entire image to give the final $g_2$ function. $\tau$ is taken to be integer multiples of the time between images. Generally, it is possible for the $g_2$ functions to have an explicit $Q$ dependence \cite{dejeu2005}, though this is not the case for this experiment - see the supplementary information for more details - and so, this is not considered for the remainder of this work. Example $g_2$ functions calculated for various points on the cooling branch measured using XMCD can be seen in Fig.~\ref{fig:x2}(f). 

To extract the dynamic behavior, the $g_2$ function is fitted by a stretched exponential model written as
\begin{equation} \label{eq:g2fit}
g_2(\tau) = 1 + A \cos (\omega \tau) e^{-\big(\frac{\tau}{\lambda}\big)^\beta},
\end{equation}
where $A$ is the speckle intensity or correlation amplitude, $\beta$ is the stretching exponent and $\lambda$ is the relaxation time. Examples of fitting this equation to the $g_2$ behaviors can be seen by the solid lines in Fig.~\ref{fig:x2}(f). The development of both $A$ and $2\pi/\omega$ with temperature can be seen in the supplementary information. As the dynamic behavior of the system is captured by the $\beta$ and $\lambda$ parameters, the extracted values of which for all temperatures for both XMCD and XMLD measurements can be seen in Fig.~\ref{fig:beta} and Fig.~\ref{fig:lambda} respectively, that is where this rest of the discussion shall focus. Measurements at some temperatures were repeated and what is shown throughout Fig.~\ref{fig:beta} and Fig.~\ref{fig:lambda} is an error weighted average of all the measurements taken at a given temperature. The fits of equation \ref{eq:g2fit} to the $g_2$ functions were performed for times where the value of the calculated $g_2$ function is larger than 1. 

Here, however, it is worth noting that this form of the fitting function is known as the heterodyne model and requires the presence of a static reference signal, for which $\omega$ represents the mixing frequency between the dynamic and static signals \cite{dejeu2005, Morley2017}. It is worth mentioning here that this model is chosen as the application of a model which does not assume the presence of a static reference signal, known as the homodyne model \cite{dejeu2005}, yields values of $\beta$ that are difficult to explain physically. More details of why this is the case will be given in the next section. The static reference signal measured here is believed to originate from the resonant charge scattering in the sample also present at the Fe L$_3$ edge.

\subsubsection{Stretching Exponent behavior}

\begin{figure}[t]
    \includegraphics[width = 8cm]{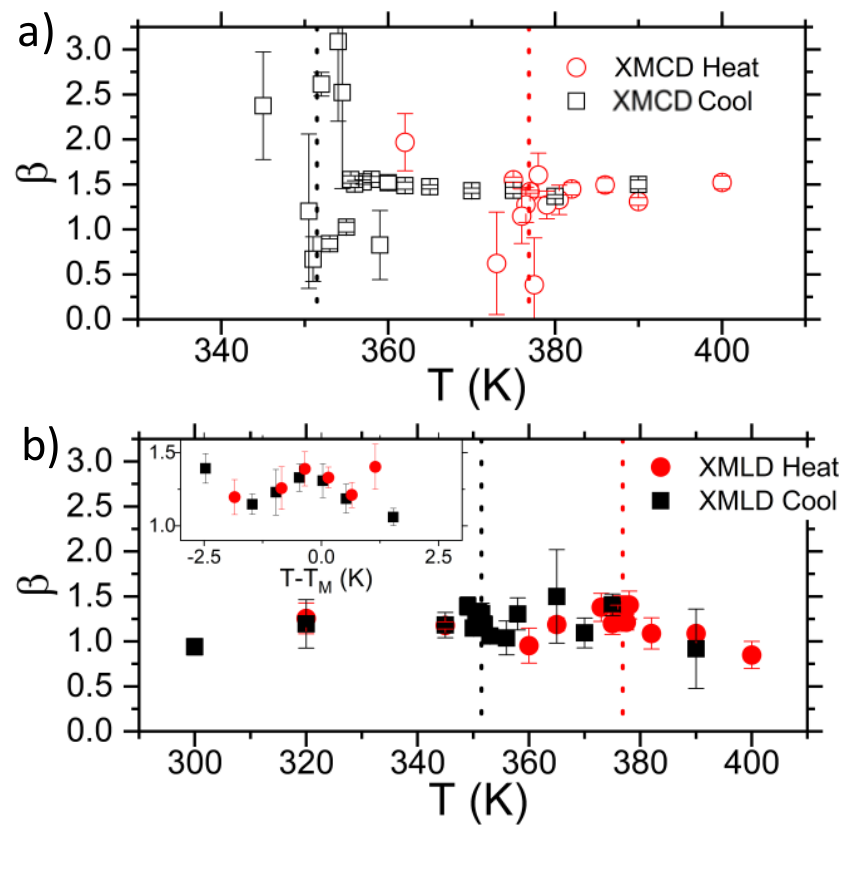}
    \caption{Stretching exponent behavior. Panels (a) and (b) show the behavior of the stretching exponent, $\beta$, extracted from the fits of equation \ref{eq:g2fit} to the $g_2$ functions against temperature, $T$, through the transition for XMCD and XMLD measurements, respectively. The inset in panel (b) shows the behavior of $\beta$ compared to the transition midpoint. The XMCD measurements show $\beta \sim 1.5$ which indicate jammed dynamic behavior, whilst the XMLD measurements appear to show $\beta$ increasing from $\sim 1 - 1.5$ close to the transition midpoint. The large scatter for measurements performed below $T_\text{M}$ when performed using XMCD is attributed to low signal in this region. 
    }
\label{fig:beta}
\end{figure}

For investigations performed using XMCD when cooling, it can be seen from Fig.~\ref{fig:beta}(a) that $\beta \approx 1.5$ for measurements performed where $T > T_\text{M}$, the transition midpoint measured from the magnetometry, which is shown by the dashed lines in this figure. For measurements performed where $T \le T_\text{M}$ the extracted values of $\beta$ develop a large scatter, meaning it is difficult to discern any coherent trends in the data. A large scatter in the data is also seen in the extracted values of $\beta$ for measurements performed for $T \le T_\text{M}$ on the heating branch of the transition. Again, for measurements performed where $T > T_\text{M}$ when heating exhibit $\beta \sim 1.5$ for all measurements. The large scatter seen in the lower temperatures for each transition branch is likely due to the low volume of FM domains within the system close to the AF phase and the resultant low signal.

The parameter $\beta$ is used to describe the nature of the statistical processes that govern the relaxation of the state \cite{Johnston2006,Hansen2013,ChenJam2013,Chen2018}. For values of $0 < \beta \le 1$ the relaxation processes are governed by thermal statistical physics \cite{Johnston2006}, meaning that the dynamic behavior can be described as diffusive \cite{ChenJam2013, Chen2018}. Whereas, values of $1 < \beta \le 2$ indicate that the relaxation is governed by Gaussian statistics \cite{Hansen2013}. In this regime the dynamics are often described as being `collective' as there are underlying long-range interactions that affect the behavior \cite{ChenJam2013, Chen2018}. Systems in which $\beta = 1.5$ exhibit what is known as `jammed' dynamics, meaning that relaxation events are unable to propagate through the system \cite{Binder1987,Johnston2006,ChenJam2013,Hansen2013,Banigan2013,Chen2018}. In magnetic systems the source of this frustration is the inability to fully resolve the Ruderman-Kittel-Kasuya-Yosida (RKKY) exchange coupling present in the system \cite{ChenJam2013}. In the case of FeRh, the inability to resolve all the exchange interactions within the region between the two magnetic phases may also contribute to this behavior \cite{Ostler2017}. It is likely that both factors contribute to the jammed behavior seen in the XMCD measurements in Fig.~\ref{fig:beta}(a).

The values of $\beta$ extracted from the fits of equation \ref{eq:g2fit} to the $g_2$ functions calculated from the XMLD measurements are shown in Fig.~\ref{fig:beta}(b). The development of $\beta$ with temperature is inherently different to that of the XMCD measurements. For temperatures in excess of $T_\text{M}$ for both transition branches, $\beta \sim 1$, which is indicative of diffusive dynamics \cite{Johnston2006, Chen2018}. When approaching $T_\text{M}$ from either direction $\beta$ increases towards $\sim 1.5$ on both transition branches as seen by the inset in Fig.~\ref{fig:beta}(b). Again, for measurements performed at temperatures lower than $T_\text{M}$ for both transition branches, $\beta$ is mostly consistent with the value 1 within an error bar.

This increase in $\beta$ when approaching $T_\text{M}$ for the XMLD measurements coincides with the introduction of FM material into the AF matrix when heating, and its removal when cooling. This change suggests that the properties of the system change fundamentally as FM material is introduced into the system. As this is a system where the AF phase is in direct contact with the FM phase, it suggests that the exchange coupling between the two magnetic phases plays a role in the dynamic behavior of the system. The introduction of exchange coupling between the AF and FM regions would introduce an extra anisotropy energy into the system which would affect the behavior of the spin-axes of both magnetic phases and the magnetic structure in the region between them. This would mean the behavior of the two magnetic phases would become intertwined leading to collective dynamics, as consistent with the experimental data. The introduction of the exchange coupling with the phase coexistence through the phase transition which changes the dynamic properties of the system when compared to the regions of the transition in which only a single phase is expected.

\subsubsection{Behavior of the Relaxation Time}

\begin{figure*}[t]
\includegraphics[width = 16cm]{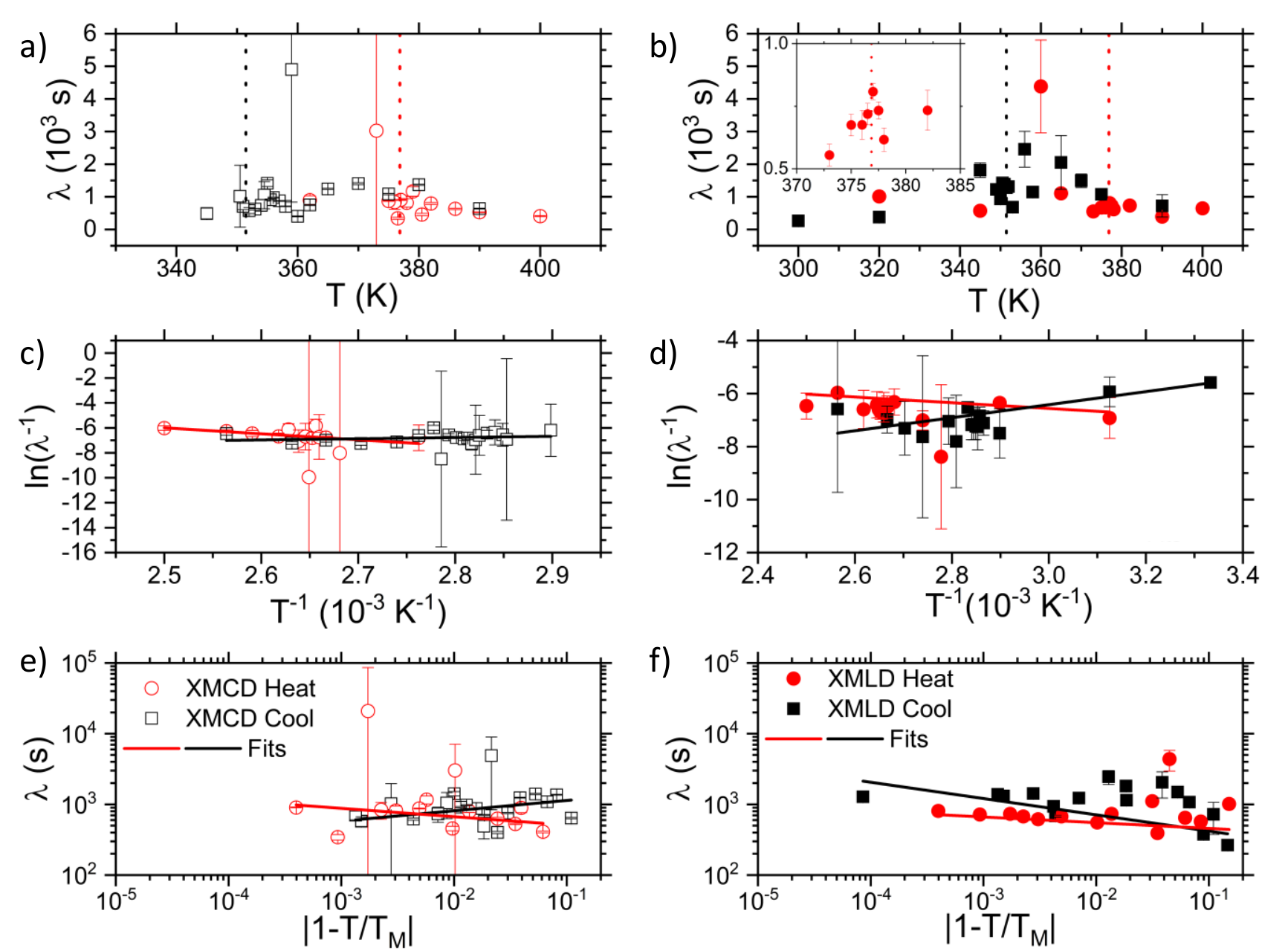}
\caption{Behavior of the relaxation time extracted from the XPCS measurements. Panels (a) and (b) show the temperature dependence of the relaxation time, $\lambda$ extracted from measurements performed using XMCD and XMLD measurements, respectively, for measurements performed both when heating and when cooling. There are no clear trend here in the data. Panels (c) and (d) show the data in panels (a) and (b) replotted in line with the Arrhenius equation seen in \ref{eq:Arr}, with the solid lines here being fits of that equation to the data. This model can be used to describe the heating branch behavior well,  but leads to a negative activation energy for measurements performed when cooling, which is a non-physical result. Panels (e) and (f) again show data in panels (a) and (b) but this time they are replotted to make it easier to fit the critical slowing down model, seen in equation \ref{eq:slowdown}, to the data. Again, the solid lines in these panels are fits of the critical slowing down model to the data. Despite the large scatter in the data, this model appears to describe the behavior of all data sets reasonably well. 
\label{fig:lambda}}
\end{figure*}

As the system jams approaching $T_\text{M}$ when measured by XMLD it may be expected that the relaxation time, $\lambda$, would increase at this stage. The values of $\lambda$ returned for the fits of Eq.~\ref{eq:g2fit} to the $g_2$ functions are shown as a function of temperature in Fig.~\ref{fig:lambda}(a) and (b) for the XMCD and XMLD experiments, respectively. $\lambda$ is typically several hundred to one thousand seconds, without any clear trend through the temperature range here.

Typically, there are three models used to explain the relaxation behaviors of magnetic systems \cite{Morley2017, Sinha2014, Djurberg1997, Chen2018}: The first is the Vogel-Fulcher-Tamann law \cite{Djurberg1997, Morley2017}, which describes the behavior of fluctuations as they approach a glass transition temperature. This model is used to describe a thermally activated process with an activation energy that varies with temperature \cite{ChenJam2013}. However, this model is asymmetric around the critical temperature which is not the case for the data here. 

The second model is that of the Arrhenius model, which is used to describe systems where a relaxation process is governed by a temperature independent activation energy, $E_\text{A}$ \cite{ChenJam2013}. The relationship between the relaxation rate $\lambda^{-1}$, where $\lambda$ is the relaxation time, and the temperature, $T$, is given by: 
\begin{equation}\label{eq:Arr}
\ln(\lambda^{-1}) = \ln(\lambda_0^{-1}) - \frac{E_\text{A}}{k_\text{B} T},
\end{equation}
where $\lambda_0^{-1}$ is the relaxation time at $T = 0$ and $k_\text{B}$ is the Boltzmann constant \cite{LovingThesis}. Arrhenius analysis was performed on the XPCS measurements and the results are shown in Fig.~\ref{fig:lambda} for measurements performed using both XMCD (panel (c)) and the XMLD (panel (d)).

The solid lines in Fig.~\ref{fig:lambda}(c) and (d) are fits of Eq.~\ref{eq:Arr} to the behavior of the relaxation time extracted from the XPCS measurements. This model is found to describe well the behavior of the measurements performed when heating, yielding $E_\text{A}/k_\text{B} = 4700 \pm 600$ K and $E_\text{A}/k_\text{B} = 1100 \pm 500$ K for the XMCD and XMLD measurements respectively. The difference here suggests an asymmetry in the nature of the behavior probed, consistent with the difference in the nature of the probe itself. This model yields a negative activation energy for both data sets taken when cooling, which is a non-physical result and implies this model cannot adequately describe the behavior seen in those datasets. Clearly, the inability of the models mentioned thus far to explain the behavior of the cooling branch behaviors requires further investigation. 

Another model used to describe relaxation behavior is that of critical slowing down, which describes the relaxation behavior for systems approaching a critical temperature associated with phase transition, $T_\text{c}$ \cite{Djurberg1997,Morley2017}.  The dependence of the relaxation time, $\lambda$, upon its proximity to $T_\text{c}$ is given by \cite{Djurberg1997,Morley2017}:
\begin{equation}
\lambda = \lambda_0 \bigg|1- \frac{T}{T_\text{c}}\bigg|^{-zv}, \label{eq:slowdown}
\end{equation}
where $zv$ is the critical scaling exponent. The fitting of this equation to the data is shown by the solid lines in Fig.~\ref{fig:lambda}(e) and (f) for the XMCD and XMLD investigations, respectively. The fits are performed using $T_\text{M}$ extracted from the magnetometry measurements. The extracted values of $zv$ from these fits are shown in Fig.~\ref{fig:zv}. The critical slowing down model is seen to describe the data reasonably well for all datasets. This implies that critical scaling of the relaxation time is observed approaching $T_\text{M}$ through the FeRh metamagnetic phase transition when probed using XPCS.

\begin{figure*}[t!]
\includegraphics[width = 16cm]{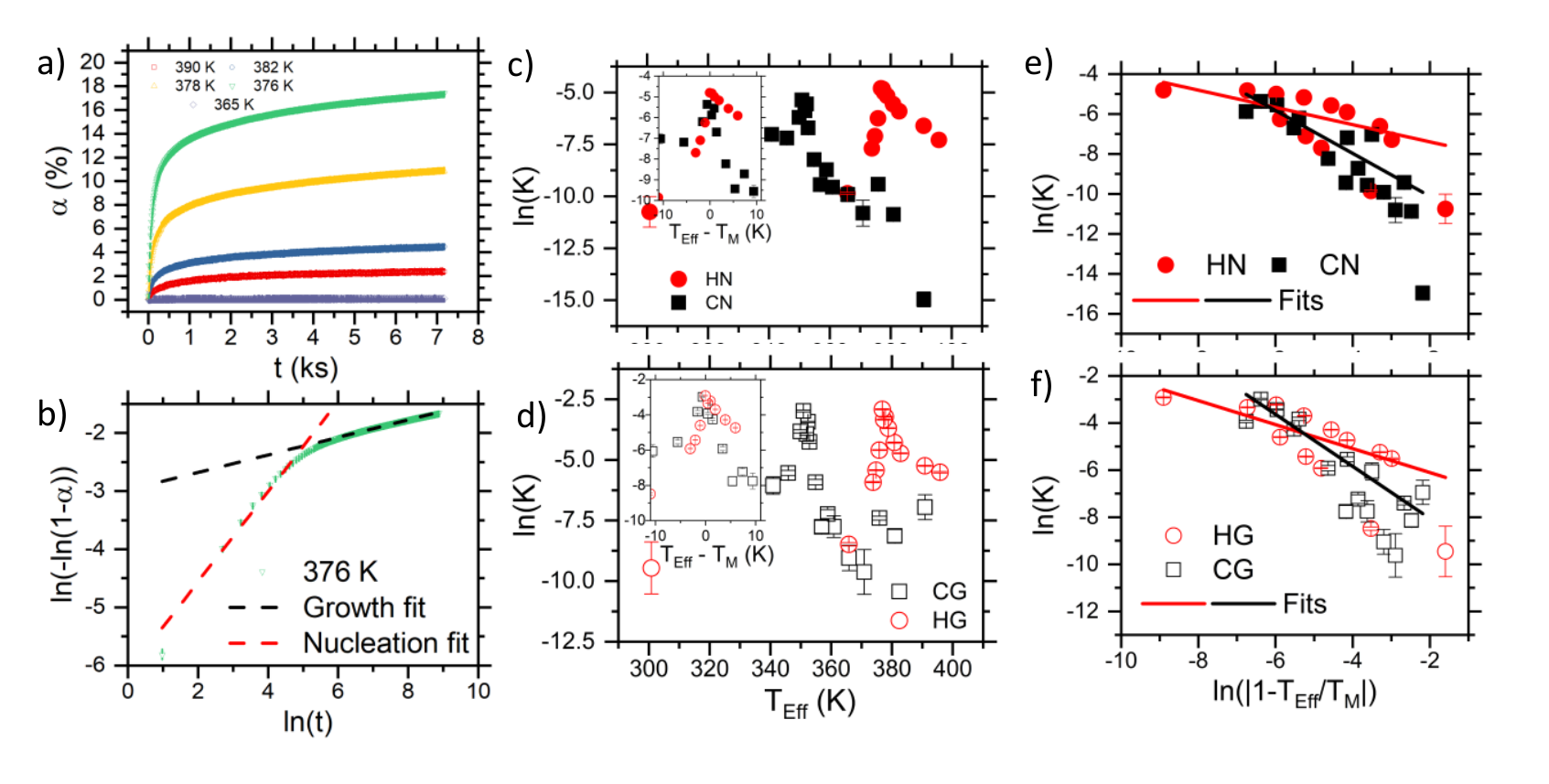}
\caption{Magnetic relaxation measured using magnetometry. (a) The evolution of $\alpha$ with time, $t$, for measurements performed at various temperatures when heating. (b) The Avrami analysis (see Methods section for description) for the measurement performed at 376 K. There are two regions where the linear relationship expected is observed here, which are believed to correspond to the nucleation (N) and subsequent growth (G) of domains in the system. The linear fits used to extract information from the two regions are shown by the dashed lines. Panels (c) and (d) show the value of $\ln (K)$ extracted from the Avrami analysis against the effective temperature, $T_\text{Eff}$, for the nucleation and growth phases respectively, for measurements performed when heating (H) and cooling (C). The insets in these figures show the behavior of $\ln(K)$ against the deviation in temperature from the transition midpoint, $T_\text{Eff} - T_\text{M}$, zoomed in around $T_\text{Eff} - T_\text{M} = 0$. These figures demonstrate an increase in $\ln (K)$ approaching $T_\text{Eff} - T_\text{M} = 0$, which is indicative of critical speeding up. Panels (e) and (f) show the behavior of $\ln (K)$ extracted for all measurements against the natural logarithm of the reduced temperature $1 - T_\text{Eff}/T_\text{M}$, for both the nucleation and growth phases respectively. The solid lines here are fits of the data to the critical slowing down model, the results of which are shown in Fig.~\ref{fig:zv}. Critical speeding up is seen for all measurements sets here. }
\label{fig:squid}
\end{figure*}

Critical scaling behavior is typically associated with second-order phase transitions where a divergence of the correlation length of thermal fluctuations is seen approaching $T_\text{c}$ \cite{Djurberg1997,Porter}. This divergence brings with it an increase in the activation energy centered around the $T_\text{c}$.The same behavior is not expected through first-order phase transitions \cite{Porter}, though critical scaling of the domain size of a given phase within the other has been seen through various first-order phase transition systems \cite{McLeod2019}, including FeRh \cite{Keavney2018}. However, in this experiment we are unable to reconcile the behavior of the relaxation time with the measured length scale (see supplemental material). 

The isothermal temporal evolution of the first-order phase transition in FeRh has previously been described using the `droplet' model \cite{LovingThesis}, which is used to model systems where one phase forms within the matrix of the other \cite{Binder1987,LovingThesis}. In this scenario, there are two main sources of fluctuations: i) the nucleation or annihilation of regions of one phase within the other and ii) fluctuations in the region between the two phases \cite{Binder1987,LovingThesis}. Either pathway could be responsible for the behavior here. So to try and isolate the source of this behavior, a similar investigation of the relaxation behavior was performed by measuring the isothermal temporal evolution of the phase transition using a SQUID-VSM, where sensitivity to the domain wall dynamics would be lost to the macroscale behavior of the domain relaxation.

\subsection{Magnetometry Measurements}
For these measurements, for comparison with the XPCS investigations, the temperature was ramped at 2 K min$^{-1}$ and the magnetization was measured for 2 hours immediately after the desired temperature was achieved. The sample was thermally cycled using the same protocol as the XPCS measurements, such that it entered the fully AF or FM state, depending on the temperature sweep direction, between measurements to reset the magnetic state The results of this study are shown in Fig.~\ref{fig:squid}.

The time dependent phase fraction, $\alpha (t)$, is defined as the ratio of the phase changed during the time interval, $t$, and the total available phase fraction available at that temperature and is calculated using\cite{LovingThesis}:
\begin{equation}
\alpha(t) = \frac{M(t) - M_\text{i}}{M_\text{S}-M_\text{i}},
\end{equation}
where $M(t)$ is the magnetization at a given time, $t$, $M_\text{i}$ is the magnetization at the beginning of the measurement time, $M_\text{S}$ is either the saturation magnetization when heating, or the residual magnetization for measurements performed when cooling.

In the Avrami model, the time dependent phase fraction can be written in terms of an exponential probability distribution with a given rate, $K$, after a time $t$ at a given temperature such that \cite{LovingThesis}:
\begin{equation}
\alpha(t) = 1 - e^{-K t^n},
\end{equation}
where $n$ is the Avrami exponent and refers to the dimensionality of the changes taking place within the system. It follows then that it is possible to extract both $K$ and $n$ using:
\begin{equation}
\ln(-\ln(1-\alpha)) = \ln(K) + n \ln(t).
\end{equation}
It also follows from this equation that regions of the data where a straight line can be used to accurately describe the behavior means that the system contains a single relaxation process with given dimensionality.

Measurements following the same protocol were also performed on a different FeRh sample, grown on MgO with the substrate still attached, through the second-order phase transition approaching the Curie temperature: see the supplementary information for more details.

The temporal evolution of the time dependent phase fraction, $\alpha(t)$, is shown for various temperatures when heating in Fig.~\ref{fig:squid}(a). $\alpha$ clearly varies through the measurement time, with the amplitude of this change peaking for measurements performed near $T_\text{M}$. By performing Avrami analysis (see Methods section) which is shown for the 376 K measurement in Fig.~\ref{fig:squid}(b), it is clear that there are two regions where the expected linear dependence on $\ln (t)$ is observed. This indicates there are two relaxation processes occurring during the measurement time: i) the initial nucleation of FM domains due to thermal equalization of the system (henceforth labelled N) and ii) the subsequent growth (G) of these domains. The value of $\ln (K)$, where $K = \lambda^{-1}$ is the rate constant, extracted from Avrami analysis is shown against $T_\text{Eff}$ for both the G and N regimes in Fig.~\ref{fig:squid}(c) and (d), respectively. All measurements here show that $\ln (K)$ increases for $T_\text{Eff} \sim T_\text{M}$ as seen in the inset where the same data is plotted against $T_\text{Eff} - T_\text{M}$, which is indicative of critical speeding up. This is confirmed by plotting the behavior of $\ln (K)$ in conjunction with the critical slowing down model described by Eq.~\ref{eq:slowdown} as seen in panels (e) and (f) of Fig.\ref{fig:squid}. The lines here represent a fit to Eq.~\ref{eq:slowdown} where the value of $T_\text{M}$ is assumed to be that extracted by magnetometry.

\begin{figure}[t]
\includegraphics[width = 8cm]{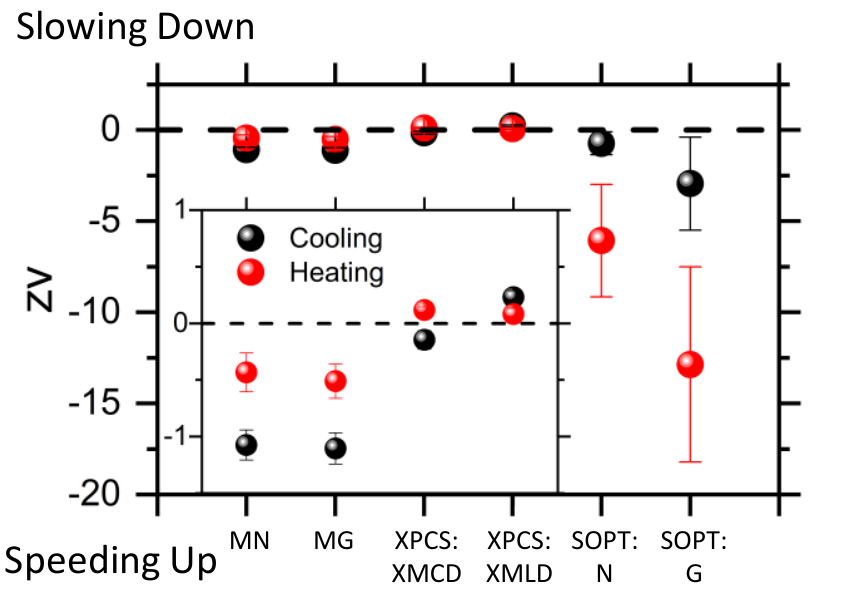}
\caption{Summary of critical scaling analysis. The value of the critical scaling exponent, $zv$, is shown for the various measurement techniques here for measurements performed both when heating and whilst cooling. Here, M means measurements performed using magnetometry through the first-order phase transition and SOPT refers to measurements performed using magnetometry through the second-order phase transition approaching the Curie temperature (see supplemental material). The definition of the critical slowing down model here means that $zv > 0$ is indicative of critical slowing down, whilst $zv < 0$ implies critical speeding up. There is a clear asymmetry between the measurements performed using magnetometry techniques and those performed using XPCS. \label{fig:zv}}
\end{figure}

Fig.~\ref{fig:zv} shows a summary of the values of $zv$ extracted for the fits of the critical slowing down model to the measurements performed using the various techniques, including through second-order phase transition at the Curie temperature (see supplemental material). Here, $zv > 0$ indicates critical slowing down, whilst critical speeding up is present for $zv < 0$. Fig.~\ref{fig:zv} clearly shows an asymmetry in the nature of the critical scaling for measurements performed using XPCS and magnetometry. 

For the measurements performed using XPCS, the extracted values of $zv$ are statistically significant from 0 within a 95\% confidence level for both the XMCD heat and XMLD cool datasets. The other two datasets have large scatter compared to their error bars and yield values of $zv$ consistent with 0 within an error bar. It is important to note that the XMCD measurements had low signal for measurements performed for $T < T_\text{M}$, which casts doubt on the reliability of this dataset. As it was also found that the Arrhenius model can be used to describe the behavior of the heating branch measurements for both dichroism types, the validity of the $zv$ result extracted for the XMCD measurements when heating is unclear. It is also worth mentioning here that despite the seeming good value extracted of $zv$ from the measurements of the cooling branch using XMLD, it is not clear again whether this model accurately describes the behaviors seen here or rather it just fits the fluctuations in the data better than the other models. Further measurements would be required to say with any certainty whether critical slowing is indeed observed in this system. What is clear from this work however, is that the behavior of the domain wall dynamics is different from that of the domain formation and growth. 

\section{Discussion}
Firstly, though there are doubts on the ability of the critical slowing down model to successfully describe the behavior of the XPCS measurements, it is clear that the dynamic behavior of the domain wall fluctuations, measured using XPCS, and the nucleation and growth of domains, measured using magnetometry techniques, is different. There is clear evidence of critical speeding up centered around the transition midpoint from the magnetometry measurements, whilst the XPCS measurements reveal a much more complicated picture.

$T_\text{M}$ is defined as the temperature where phase coexistence is maximised and where the difference in the free energy of the two states is minimised. The rate of domain nucleation would increase for $T \sim T_\text{M}$, meaning the critical speeding up seen in the magnetometry measurements can be accounted for easily. It would be expected that the behavior of the domain wall fluctuations would follow the same temperature dependence, where instead the Arrhenius law can be used to describe the behavior seen when heating, and none of the models used in this work can describe the behavior of the cooling branch measurements effectively. We can say, however, that the dynamic behavior of the domain walls when measured using XMLD is different in the region where there is phase coexistence to the behavior seen in the nominally AF or FM phase, as evidenced by the change in $\beta$ for $T \sim T_\text{M}$. This change in behavior coincides with the introduction of phase coexistence into the system. Implying that the introduction of the exchange coupling into the system has a profound affect on the dynamic behavior of the systems' domain walls. 

Previous studies into the nature of the exchange coupling in this material has been seen to develop through the phase transition in a manner consistent with thickness dependent phase transitions in AF/FM bilayer systems \cite{Massey2019}. The coupling between phases acts to overcome the exchange energy in the FM regions, causing a blurring of the magnetic order in the region between the two phases \cite{Massey2019}. It may be then that this behavior can be attributed to the region between the two phases, the phase boundary wall. This structure has been predicted to be an agglomeration of regions of both magnetic phases with dimensions between 1 - 5 nm \cite{Massey2019}. The size of this object mean it is not possible to observe its influence directly, as it falls outside the available $Q$ range, but we believe that what we are seeing in this work is the influence of the phase boundary wall has on other objects within the system, though further work is required to state this with any certainty. The exact role of this coupling in the dynamic behavior is unclear at this stage, but it may act against the influence of latent heat in the region where the coupling is strongest, which would give the phase boundary wall different properties to the bulk regions of either magnetic phase.

Little is known about the region between phases in a first-order phase transition system but it is clear that it plays a role in the dynamic behavior. Given there are many ways that two states in a first-order phase transition system can couple across the boundary including elastic coupling due to strain \cite{Lee2015, Kundu2020}, magnetic coupling in metamagnetic transitions \cite{Massey2019, Gray2019}, and charge coupling in metal-insulator systems \cite{McLeod2019, Post2018}, it may be that this influence of the coupling across the phase boundary can be seen to affect a variety of different first-order phase transition systems. This work highlights that the influence of interphase coupling should be considered in theories of first-order phase transition dynamics when interested in the smaller scale behavior, as well as demonstrating that the phase boundary wall to be an interesting entity in its own right which requires further study. We hope, as the next generation of synchrotron sources become available it will be possible to resolve the structure of the phase boundary wall in this, and other, first-order phase transition systems. 

\section{Conclusion}

To conclude, XPCS investigations were performed using both XMCD and XMLD to measure the dynamic behavior of the system through the FeRh metamagnetic phase transition. When probed using XMCD, investigations of the static structure of the scattering ring reveal a change in the nature of the scatterer when heating through the transition from domains to domain walls and back again when cooling. Whereas, measurements performed using XMLD show a change in the nature of the scatterer from being the FM material to AF material centered around the transition midpoint. 

The dynamic behavior extracted from the XPCS measurements paints a much more complicated picture. Measurements performed using XMCD reveal jammed dynamics where $T > T_text{M}$ and a large scatter below that, most likely due to the low volume of FM material at low temperatures. The nature of the dynamic behavior appears to undergo a transition from diffusive to jammed dynamics centered around $T \sim t_\text{M}$. This is attributed to the exchange coupling that accompanies the introduction of phase coexistence into the system as the transition progresses. It is found that the Arrhenius model can be used to describe the behavior of the relaxation time of the measurements performed when heating. Whereas, the behavior of the relaxation time for the cooling branch measurements cannot be described well with any of the Arrhenius, critical slowing down or Vogel-Fulcher-Tammann models. 

To try and ascertain the source of these behaviors, measurements of the dynamic behavior on the macroscale were performed using magnetometry techniques. These investigations reveal critical speeding up around the transition midpoint, which can be attributed to the reduction in the free energy difference between the two states at this point. The macroscale investigations imply that the behavior measured using XPCS belongs to domain wall fluctuations, where the influence of the interphase coupling on the behavior of the relaxation behavior can be clearly seen in the XMLD measurements. The exchange coupling changes the dynamics of domain walls where there is phase coexistence compared to the domain wall behavior in either the nominally AF or FM phase. We believe that this change in the behavior comes from the phase boundary wall, whose influence we see in the behavior of other objects as it is too small to be measured directly in this experiment. This work highlights the importance of considering the role of the interphase coupling in models of first-order phase transition dynamics on the microscale. 

\begin{acknowledgments}
This work was supported by the Diamond Light Source and the UK EPSRC (grant numbers EP/M018504/1 and EP/M019020/1).
\end{acknowledgments}

\bibliographystyle{naturemag}

%\bibliography{Massey_thesis}

\begin{thebibliography}{10}
\expandafter\ifx\csname url\endcsname\relax
  \def\url#1{\texttt{#1}}\fi
\expandafter\ifx\csname urlprefix\endcsname\relax\def\urlprefix{URL }\fi
\providecommand{\bibinfo}[2]{#2}
\providecommand{\eprint}[2][]{\url{#2}}

\bibitem{Porter}
\bibinfo{author}{Porter, D.~A.} \& \bibinfo{author}{Easterling, K.~E.}
\newblock \emph{\bibinfo{title}{Phase Transformations in Metals and Alloys}},
  chap.~\bibinfo{chapter}{5} (\bibinfo{publisher}{Van Nostrand Reinhold
  Publishing}, \bibinfo{year}{1981}).

\bibitem{Djurberg1997}
\bibinfo{author}{Djurberg, C.} \emph{et~al.}
\newblock \bibinfo{title}{Dynamics of an interacting particle system: Evidence
  of critical slowing down}.
\newblock \emph{\bibinfo{journal}{Phys. Rev. Lett.}}
  \textbf{\bibinfo{volume}{79}}, \bibinfo{pages}{5154} (\bibinfo{year}{1997}).

\bibitem{Morley2017}
\bibinfo{author}{Morley, S.~A.} \emph{et~al.}
\newblock \bibinfo{title}{{Vogel-Fulcher-Tammann} freezing of a thermally
  fluctuating artificial spin ice probed by x-ray photon correlation
  spectroscopy}.
\newblock \emph{\bibinfo{journal}{Phys. Rev. B}} \textbf{\bibinfo{volume}{95}},
  \bibinfo{pages}{104422} (\bibinfo{year}{2017}).

\bibitem{McLeod2019}
\bibinfo{author}{McLeod, A.~S.} \emph{et~al.}
\newblock \bibinfo{title}{Nanotextured phase coexistence in the correlated
  insulator {V$_2$O$_3$}}.
\newblock \emph{\bibinfo{journal}{Nat. Physics.}}
  \textbf{\bibinfo{volume}{13}}, \bibinfo{pages}{80} (\bibinfo{year}{2016}).

\bibitem{Nemoto2017}
\bibinfo{author}{Nemoto, T.}, \bibinfo{author}{Jack, R.~L.} \&
  \bibinfo{author}{Lecomte, V.}
\newblock \bibinfo{title}{Finite-size scaling of a first-order dynamical phase
  transition: Adaptive population dynamics and an effective model}.
\newblock \emph{\bibinfo{journal}{Phys. Rev. Lett.}}
  \textbf{\bibinfo{volume}{118}}, \bibinfo{pages}{115702}
  (\bibinfo{year}{2017}).

\bibitem{Post2018}
\bibinfo{author}{Post, K.~W.} \emph{et~al.}
\newblock \bibinfo{title}{Coexisting first- and second-order electronic phase
  transitions in a correlated oxide}.
\newblock \emph{\bibinfo{journal}{Nat. Physics.}}
  \textbf{\bibinfo{volume}{14}}, \bibinfo{pages}{1056} (\bibinfo{year}{2018}).

\bibitem{Keavney2018}
\bibinfo{author}{Keavney, D.~J.} \emph{et~al.}
\newblock \bibinfo{title}{Phase coexistence and kinetic arrest in the
  magnetostructural transition of the ordered alloy {FeRh}}.
\newblock \emph{\bibinfo{journal}{Sci. Rep.}} \textbf{\bibinfo{volume}{8}},
  \bibinfo{pages}{1778} (\bibinfo{year}{2018}).

\bibitem{Kundu2020}
\bibinfo{author}{Kundu, S.}, \bibinfo{author}{Bar, T.}, \bibinfo{author}{Nayak,
  R.~K.} \& \bibinfo{author}{Bansal, B.}
\newblock \bibinfo{title}{Critical slowing down at the abrupt mott transition:
  When the first-order phase transition becomes zeroth order and looks like
  second order}.
\newblock \emph{\bibinfo{journal}{Phys. Rev. Lett.}}
  \textbf{\bibinfo{volume}{124}}, \bibinfo{pages}{095703}
  (\bibinfo{year}{2020}).

\bibitem{Chen2018}
\bibinfo{author}{Chen, X.~M.} \emph{et~al.}
\newblock \bibinfo{title}{Spontaneous magnetic superdomain wall fluctuations in
  an artificial antiferromagnet}.
\newblock \emph{\bibinfo{journal}{Phys. Rev. Lett.}}
  \textbf{\bibinfo{volume}{123}}, \bibinfo{pages}{197202}
  (\bibinfo{year}{2019}).

\bibitem{Maat2005}
\bibinfo{author}{Maat, S.}, \bibinfo{author}{Thiele, J.~U.} \&
  \bibinfo{author}{Fullerton, E.~E.}
\newblock \bibinfo{title}{Temperature and field hysteresis of the
  antiferromagnetic-to-ferromagnetic phase transition in epitaxial {FeRh}
  films}.
\newblock \emph{\bibinfo{journal}{Phys. Rev. B}} \textbf{\bibinfo{volume}{72}},
  \bibinfo{pages}{214432} (\bibinfo{year}{2005}).

\bibitem{Moriyama2015}
\bibinfo{author}{Moriyama, T.} \emph{et~al.}
\newblock \bibinfo{title}{Sequential write-read operations in {FeRh}
  antiferromagnetic memory}.
\newblock \emph{\bibinfo{journal}{Appl. Phys. Lett.}}
  \textbf{\bibinfo{volume}{107}}, \bibinfo{pages}{122403}
  (\bibinfo{year}{2015}).

\bibitem{Sinha2014}
\bibinfo{author}{Sinha, S.~K.}, \bibinfo{author}{Jiang, Z.} \&
  \bibinfo{author}{Lurio, L.~B.}
\newblock \bibinfo{title}{X-ray photon correlated spectroscopy studies of
  surfaces and thin films}.
\newblock \emph{\bibinfo{journal}{Advanced Materials}}
  \textbf{\bibinfo{volume}{26}}, \bibinfo{pages}{7764} (\bibinfo{year}{2014}).

\bibitem{Shpyrko2007}
\bibinfo{author}{Shpyrko, O.~G.} \emph{et~al.}
\newblock \bibinfo{title}{Direct measurement of antiferromagnetic domain
  fluctuations}.
\newblock \emph{\bibinfo{journal}{Nature}} \textbf{\bibinfo{volume}{447}},
  \bibinfo{pages}{68} (\bibinfo{year}{2007}).

\bibitem{Konings2011}
\bibinfo{author}{Konings, S.} \emph{et~al.}
\newblock \bibinfo{title}{Magnetic domain fluctuations in an antiferromagnetic
  film observed with coherent resonant soft x-ray scattering}.
\newblock \emph{\bibinfo{journal}{Physical Review Letters}}
  \textbf{\bibinfo{volume}{106}}, \bibinfo{pages}{077402}
  (\bibinfo{year}{2011}).

\bibitem{ChenJam2013}
\bibinfo{author}{Chen, S.-W.} \emph{et~al.}
\newblock \bibinfo{title}{Jamming behavior of domains in a spiral
  antiferromagnetic system}.
\newblock \emph{\bibinfo{journal}{Phys. Rev. Lett.}}
  \textbf{\bibinfo{volume}{110}}, \bibinfo{pages}{217201}
  (\bibinfo{year}{2013}).

\bibitem{Swart1984}
\bibinfo{author}{Swartzendruber, L.}
\newblock \bibinfo{title}{The {Fe-Rh} (iron-rhodium) system}.
\newblock \emph{\bibinfo{journal}{Journal of Phase Equilibria}}
  \textbf{\bibinfo{volume}{5}}, \bibinfo{pages}{456} (\bibinfo{year}{1984}).

\bibitem{Fallot1939}
\bibinfo{author}{Fallot, M.} \& \bibinfo{author}{Hocart, R.}
\newblock \bibinfo{title}{On the appearance of ferromagnetism upon elevation of
  the temperature of iron and rhodium}.
\newblock \emph{\bibinfo{journal}{Rev. Sci.}} \textbf{\bibinfo{volume}{8}},
  \bibinfo{pages}{498} (\bibinfo{year}{1939}).

\bibitem{Kouvel1962}
\bibinfo{author}{Kouvel, J.~S.} \& \bibinfo{author}{Hartelius, C.~C.}
\newblock \bibinfo{title}{Anomalous magnetic moments and transformations in the
  ordered alloy {FeRh}}.
\newblock \emph{\bibinfo{journal}{J. Appl. Phys.}}
  \textbf{\bibinfo{volume}{33}}, \bibinfo{pages}{1343} (\bibinfo{year}{1962}).

\bibitem{Kouvel1966}
\bibinfo{author}{Kouvel, J.~S.}
\newblock \bibinfo{title}{Unusual nature of the abrupt magnetic transition in
  {FeRh} and its pseudobinary variants}.
\newblock \emph{\bibinfo{journal}{J. Appl. Phys.}}
  \textbf{\bibinfo{volume}{37}}, \bibinfo{pages}{1257} (\bibinfo{year}{1966}).

\bibitem{devries2013}
\bibinfo{author}{de~Vries, M.~A.} \emph{et~al.}
\newblock \bibinfo{title}{Hall-effect characterization of the metamagnetic
  transition in {FeRh}}.
\newblock \emph{\bibinfo{journal}{N. J. Phys.}} \textbf{\bibinfo{volume}{15}},
  \bibinfo{pages}{013008} (\bibinfo{year}{2013}).

\bibitem{Ricodeau1972}
\bibinfo{author}{Ricodeau, J.~A.} \& \bibinfo{author}{Melville, D.}
\newblock \bibinfo{title}{Model of the antiferromagnetic-ferromagnetic
  transition in {FeRh} alloys}.
\newblock \emph{\bibinfo{journal}{Journal of Physics F: Metal Physics}}
  \textbf{\bibinfo{volume}{2}}, \bibinfo{pages}{337} (\bibinfo{year}{1972}).

\bibitem{Thiele2004}
\bibinfo{author}{Thiele, J.-U.}, \bibinfo{author}{Buess, M.} \&
  \bibinfo{author}{Back, C.~H.}
\newblock \bibinfo{title}{Spin dynamics of the
  antiferromagnetic-to-ferromagnetic phase transition in {FeRh} on a
  sub-picosecond time scale}.
\newblock \emph{\bibinfo{journal}{Appl. Phys. Lett.}}
  \textbf{\bibinfo{volume}{85}}, \bibinfo{pages}{2857} (\bibinfo{year}{2004}).

\bibitem{Ju2004}
\bibinfo{author}{Ju, G.} \emph{et~al.}
\newblock \bibinfo{title}{Ultrafast generation of ferromagnetic order via a
  laser-induced phase transformation in {FeRh} thin films}.
\newblock \emph{\bibinfo{journal}{Phys. Rev. Lett.}}
  \textbf{\bibinfo{volume}{93}}, \bibinfo{pages}{197403}
  (\bibinfo{year}{2004}).

\bibitem{Radu2010}
\bibinfo{author}{Radu, I.} \emph{et~al.}
\newblock \bibinfo{title}{Laser-induced generation and quenching of
  magnetization on {FeRh} studied with time-resolved x-ray magnetic circular
  dichroism}.
\newblock \emph{\bibinfo{journal}{Phys. Rev. B}} \textbf{\bibinfo{volume}{81}},
  \bibinfo{pages}{104415} (\bibinfo{year}{2010}).

\bibitem{Thiele2003}
\bibinfo{author}{Thiele, J.~U.}, \bibinfo{author}{Maat, S.} \&
  \bibinfo{author}{Fullerton, E.~E.}
\newblock \bibinfo{title}{{FeRh/FePt} exchange spring films for thermally
  assisted magnetic recording media}.
\newblock \emph{\bibinfo{journal}{Appl. Phys. Lett.}}
  \textbf{\bibinfo{volume}{82}}, \bibinfo{pages}{2859} (\bibinfo{year}{2003}).

\bibitem{Cherifi2014}
\bibinfo{author}{Cherifi, R.~O.} \emph{et~al.}
\newblock \bibinfo{title}{Electric-field control of magnetic order above room
  temperature}.
\newblock \emph{\bibinfo{journal}{Nat. Mater.}} \textbf{\bibinfo{volume}{13}},
  \bibinfo{pages}{345} (\bibinfo{year}{2014}).

\bibitem{Lee2015}
\bibinfo{author}{Lee, Y.} \emph{et~al.}
\newblock \bibinfo{title}{Large resistivity modulation in mixed-phase metallic
  systems}.
\newblock \emph{\bibinfo{journal}{Nat. Comms.}} \textbf{\bibinfo{volume}{6}},
  \bibinfo{pages}{5959} (\bibinfo{year}{2015}).

\bibitem{Marti2014}
\bibinfo{author}{Marti, X.} \emph{et~al.}
\newblock \bibinfo{title}{Room-temperature antiferromagnetic memory resistor}.
\newblock \emph{\bibinfo{journal}{Nat. Mater.}} \textbf{\bibinfo{volume}{13}},
  \bibinfo{pages}{367} (\bibinfo{year}{2014}).

\bibitem{LeGraet2015}
\bibinfo{author}{Le~Gra\"{e}t, C.} \emph{et~al.}
\newblock \bibinfo{title}{Temperature controlled motion of an antiferromagnet-
  ferromagnet interface within a dopant-graded {FeRh} epilayer}.
\newblock \emph{\bibinfo{journal}{APL Materials}} \textbf{\bibinfo{volume}{3}},
  \bibinfo{pages}{041802} (\bibinfo{year}{2015}).

\bibitem{Temple2019}
\bibinfo{author}{Temple, R.~C.} \emph{et~al.}
\newblock \bibinfo{title}{Phase domain boundary motion and memristance in
  gradient-doped {FeRh} nanopillars induced by spin injection}.
\newblock \emph{\bibinfo{journal}{arXiv e-prints}}
  \bibinfo{pages}{arXiv:1905.03573} (\bibinfo{year}{2019}).

\bibitem{Kande2011}
\bibinfo{author}{Kande, D.}, \bibinfo{author}{Pisana, S.},
  \bibinfo{author}{Weller, D.}, \bibinfo{author}{Laughlin, D.~E.} \&
  \bibinfo{author}{Zhu, J.~G.}
\newblock \bibinfo{title}{Enhanced {B2} ordering of {FeRh} thin films using {B2
  NiAl} underlayers}.
\newblock \emph{\bibinfo{journal}{IEEE Transactions on Magnetics}}
  \textbf{\bibinfo{volume}{47}}, \bibinfo{pages}{3296} (\bibinfo{year}{2011}).

\bibitem{Massey2019}
\bibinfo{author}{Massey, J.~R.} \emph{et~al.}
\newblock \bibinfo{title}{Phase boundary exchange coupling in the mixed
  magnetic phase regime of a pd-doped ferh epilayer}.
\newblock \emph{\bibinfo{journal}{Phys. Rev. Materials}}
  \textbf{\bibinfo{volume}{4}}, \bibinfo{pages}{024403} (\bibinfo{year}{2020}).

\bibitem{Almeida2017}
\bibinfo{author}{Almeida, T.~P.} \emph{et~al.}
\newblock \bibinfo{title}{Quantitative {TEM} imaging of the magnetostructural
  and phase transitions in {FeRh} thin film systems}.
\newblock \emph{\bibinfo{journal}{Sci. Rep.}} \textbf{\bibinfo{volume}{7}},
  \bibinfo{pages}{17835} (\bibinfo{year}{2017}).

\bibitem{Temple2018}
\bibinfo{author}{Temple, R.~C.} \emph{et~al.}
\newblock \bibinfo{title}{Antiferromagnetic-ferromagnetic phase domain
  development in nanopatterned {FeRh} islands}.
\newblock \emph{\bibinfo{journal}{Phys. Rev. Materials}}
  \textbf{\bibinfo{volume}{2}}, \bibinfo{pages}{104406} (\bibinfo{year}{2018}).

\bibitem{Russell2013}
\bibinfo{author}{Russell, C.} \emph{et~al.}
\newblock \bibinfo{title}{Spectroscopy of polycrystalline materials using
  thinned-substrate planar goubau line at cryogenic temperatures}.
\newblock \emph{\bibinfo{journal}{Lab on a Chip}}
  \textbf{\bibinfo{volume}{13}}, \bibinfo{pages}{4065} (\bibinfo{year}{2013}).

\bibitem{Fischer1997}
\bibinfo{author}{Fischer, P.}, \bibinfo{author}{Zeller, R.},
  \bibinfo{author}{Sch\"utz, G.}, \bibinfo{author}{Georigk, G.} \&
  \bibinfo{author}{Haubold, H.}
\newblock \bibinfo{title}{Magnetic small angle x-ray scattering}.
\newblock \emph{\bibinfo{journal}{Journal de Physique {IV}}}
  \textbf{\bibinfo{volume}{7}}, \bibinfo{pages}{753} (\bibinfo{year}{1997}).

\bibitem{Sivia}
\bibinfo{author}{Sivia, D.~S.}
\newblock \emph{\bibinfo{title}{Elementary Scattering Theory For X-Ray and
  Neutron Users}}, chap.~\bibinfo{chapter}{5} (\bibinfo{publisher}{Oxford
  University Press}, \bibinfo{year}{2011}).

\bibitem{Bagschik2016}
\bibinfo{author}{Bagschik, K.} \emph{et~al.}
\newblock \bibinfo{title}{Employing soft x-ray resonant magnetic scattering to
  study domain sizes and anisotropy in {Co/Pd} multilayers}.
\newblock \emph{\bibinfo{journal}{Phys. Rev. B}} \textbf{\bibinfo{volume}{94}},
  \bibinfo{pages}{134413} (\bibinfo{year}{2016}).

\bibitem{stohrbook}
\bibinfo{author}{Stohr, J.} \& \bibinfo{author}{Siegmann, H.~C.}
\newblock \emph{\bibinfo{title}{Magnetism: From Fundamentals to Nanoscale
  Dynamics}}, chap.~\bibinfo{chapter}{9} (\bibinfo{publisher}{Springer Series
  in Solid State Physics}, \bibinfo{year}{2006}).

\bibitem{Baldasseroni2012}
\bibinfo{author}{Baldasseroni, C.} \emph{et~al.}
\newblock \bibinfo{title}{Temperature-driven nucleation of ferromagnetic
  domains in {FeRh} thin films}.
\newblock \emph{\bibinfo{journal}{Appl. Phys. Lett.}}
  \textbf{\bibinfo{volume}{100}}, \bibinfo{pages}{262401}
  (\bibinfo{year}{2012}).

\bibitem{Kinane2014}
\bibinfo{author}{Kinane, C.~J.} \emph{et~al.}
\newblock \bibinfo{title}{Observation of a temperature dependent asymmetry in
  the domain structure of a {Pd-doped FeRh} epilayer}.
\newblock \emph{\bibinfo{journal}{N. J. Phys.}} \textbf{\bibinfo{volume}{16}},
  \bibinfo{pages}{113073} (\bibinfo{year}{2014}).

\bibitem{Mariager2013}
\bibinfo{author}{Mariager, S.~O.}, \bibinfo{author}{Le~Guyader, L.},
  \bibinfo{author}{Buzzi, M.}, \bibinfo{author}{Ingold, G.} \&
  \bibinfo{author}{Quitmann, C.}
\newblock \bibinfo{title}{Imaging the antiferromagnetic to ferromagnetic first
  order phase transition of {FeRh}}.
\newblock \emph{\bibinfo{journal}{arXiv e-prints}}
  \bibinfo{pages}{arXiv:1301.4164} (\bibinfo{year}{2013}).

\bibitem{Baldasseroni2015}
\bibinfo{author}{Baldasseroni, C.} \emph{et~al.}
\newblock \bibinfo{title}{Temperature-driven growth of antiferromagnetic
  domains in thin-film {FeRh}}.
\newblock \emph{\bibinfo{journal}{J. Phys. : Cond. Mater.}}
  \textbf{\bibinfo{volume}{27}}, \bibinfo{pages}{256001}
  (\bibinfo{year}{2015}).

\bibitem{dejeu2005}
\bibinfo{author}{de~Jeu, W.~H.}, \bibinfo{author}{Madsen, A.},
  \bibinfo{author}{Sikharulidze, I.} \& \bibinfo{author}{Sprunt, S.}
\newblock \bibinfo{title}{Heterodyne and homodyne detection in fluctuating
  smectic membranes by photon correlation spectroscopy at x-ray and visible
  wavelengths}.
\newblock \emph{\bibinfo{journal}{Physica B: Cond. Mat.}}
  \textbf{\bibinfo{volume}{357}}, \bibinfo{pages}{39} (\bibinfo{year}{2005}).

\bibitem{Johnston2006}
\bibinfo{author}{Johnston, D.~C.}
\newblock \bibinfo{title}{Stretched exponential relaxation arising from a
  continuous sum of exponential decays}.
\newblock \emph{\bibinfo{journal}{Phys. Rev. B}} \textbf{\bibinfo{volume}{74}},
  \bibinfo{pages}{184430} (\bibinfo{year}{2006}).

\bibitem{Hansen2013}
\bibinfo{author}{Hansen, E.~W.}, \bibinfo{author}{Gong, X.} \&
  \bibinfo{author}{Chen, Q.}
\newblock \bibinfo{title}{Compressed exponential response function arising from
  a continuous distribution of gaussian decays - distribution characteristics}.
\newblock \emph{\bibinfo{journal}{Macro. Chem. and Phys.}}
  \textbf{\bibinfo{volume}{214}}, \bibinfo{pages}{844} (\bibinfo{year}{2013}).

\bibitem{Binder1987}
\bibinfo{author}{Binder, K.}
\newblock \bibinfo{title}{Theory of first-order phase transitions}.
\newblock \emph{\bibinfo{journal}{Reports on Progress in Physics}}
  \textbf{\bibinfo{volume}{50}}, \bibinfo{pages}{783} (\bibinfo{year}{1987}).

\bibitem{Banigan2013}
\bibinfo{author}{Banigan, E.~J.}, \bibinfo{author}{Illich, M.~K.},
  \bibinfo{author}{Stace-Naughton, D.~J.} \& \bibinfo{author}{Egolf, D.~A.}
\newblock \bibinfo{title}{The chaotic dynamics of jamming}.
\newblock \emph{\bibinfo{journal}{Nature Physics}}
  \textbf{\bibinfo{volume}{9}}, \bibinfo{pages}{288} (\bibinfo{year}{2013}).

\bibitem{Ostler2017}
\bibinfo{author}{Ostler, T.~A.}, \bibinfo{author}{Barton, C.},
  \bibinfo{author}{Thomson, T.} \& \bibinfo{author}{Hrkac, G.}
\newblock \bibinfo{title}{Modeling the thickness dependence of the magnetic
  phase transition temperature in thin {FeRh} films}.
\newblock \emph{\bibinfo{journal}{Phys. Rev. B}} \textbf{\bibinfo{volume}{95}},
  \bibinfo{pages}{064415} (\bibinfo{year}{2017}).

\bibitem{LovingThesis}
\bibinfo{author}{Loving, M.}
\newblock \emph{\bibinfo{title}{Understanding the Magnetostructural
  Transformation in {FeRh} thin films}}.
\newblock Ph.D. thesis, \bibinfo{school}{The Department of Chemical
  Engineering, Northeastern University} (\bibinfo{year}{2013}).

\bibitem{Gray2019}
\bibinfo{author}{Gray, I.} \emph{et~al.}
\newblock \bibinfo{title}{Imaging uncompensated moments and exchange-biased
  emergent ferromagnetism in {FeRh} thin films}.
\newblock \emph{\bibinfo{journal}{Phys. Rev. Materials}}
  \textbf{\bibinfo{volume}{3}}, \bibinfo{pages}{124407} (\bibinfo{year}{2019}).

\end{thebibliography}

%\input{Main_Paper.bbl}

\end{document}

%\preprint{APS/123-QED}

\title{Asymmetric Magnetic Relaxation Behaviour of Domains and Domain Walls Observed Through the FeRh First-Order Metamagnetic Phase Transition}% Force line breaks with \\
\author{Jamie R.~Massey}
\email{jamie.massey@psi.ch}
\affiliation{School of Physics and Astronomy, University of Leeds, Leeds LS2 9JT, United Kingdom.}

\author{Rowan C.~Temple}
\affiliation{School of Physics and Astronomy, University of Leeds, Leeds LS2 9JT, United Kingdom.}

\author{Trevor P.~Almeida}
\affiliation{SUPA, School of Physics and Astronomy, University of Glasgow, Glasgow, G12 8QQ, United Kingdom.}

\author{Ray Lamb}
\affiliation{School of Physics and Astronomy, University of Glasgow, Glasgow, G12 8QQ, United Kingdom.}

\author{Nicolas A.~Peters}
\affiliation{School of Physics and Astronomy, University of Leeds, Leeds LS2 9JT, United Kingdom.}
\affiliation{School of Electronic and Electrical Engineering, University of Leeds, Leeds, LS2 9JT, United Kingdom.}

\author{Richard P.~Campion}
\affiliation{School of Physics and Astronomy, University of Nottingham, Nottingham, NG7 2RD, United Kingdom.}

\author{Raymond Fan}
\affiliation{Diamond Light Source, Chilton, Didcot, OX11 0DE, United Kingdom.}

\author{Damien McGrouther}
\affiliation{SUPA, School of Physics and Astronomy, University of Glasgow, Glasgow, G12 8QQ, United Kingdom.}

\author{Stephen McVitie}
\affiliation{SUPA, School of Physics and Astronomy, University of Glasgow, Glasgow, G12 8QQ, United Kingdom.}

\author{Paul Steadman}
\affiliation{Diamond Light Source, Chilton, Didcot, OX11 0DE, United Kingdom.}

\author{Christopher H.~Marrows}
\email{c.h.marrows@leeds.ac.uk}
\affiliation{School of Physics and Astronomy, University of Leeds, Leeds LS2 9JT, United Kingdom.}


\maketitle
\section{Effective Temperature}
The transition temperature, $T_\text{T}$, of the metamagnetic phase transition in FeRh is sensitive to the application of external magnetic field, $H_\text{Ext}$ \cite{Maat2005, Massey2019}. As such, when comparing between measurement performed whilst a magnetic field is applied to those performed in the absence of an external magnetic field, it is necessary to correct the temperature of the measurements in which a field is applied. This corrected temperature is known as the effective temperature, $T_\text{Eff}$, and is defined as the temperature at which the same magnetization is expected in the absence of external field \cite{Massey2019}. The formula used for its calculation is given by,
\begin{equation}
    T_\text{Eff} = T_0 - \frac{dT_\text{T}}{d(\mu_0 H_\text{Ext})}\mu_0 H_\text{Ext},
\end{equation}
where $T_0$ is the real sample temperature, $\mu_0$ is the permeability of free space and  $dT_\text{T}/d(\mu_0 H_\text{Ext}$ is the rate of change of the $T_\text{T}$ with $\mu_0 H_\text{Ext}$.

\begin{figure*}[t]
    \includegraphics[width = 16cm]{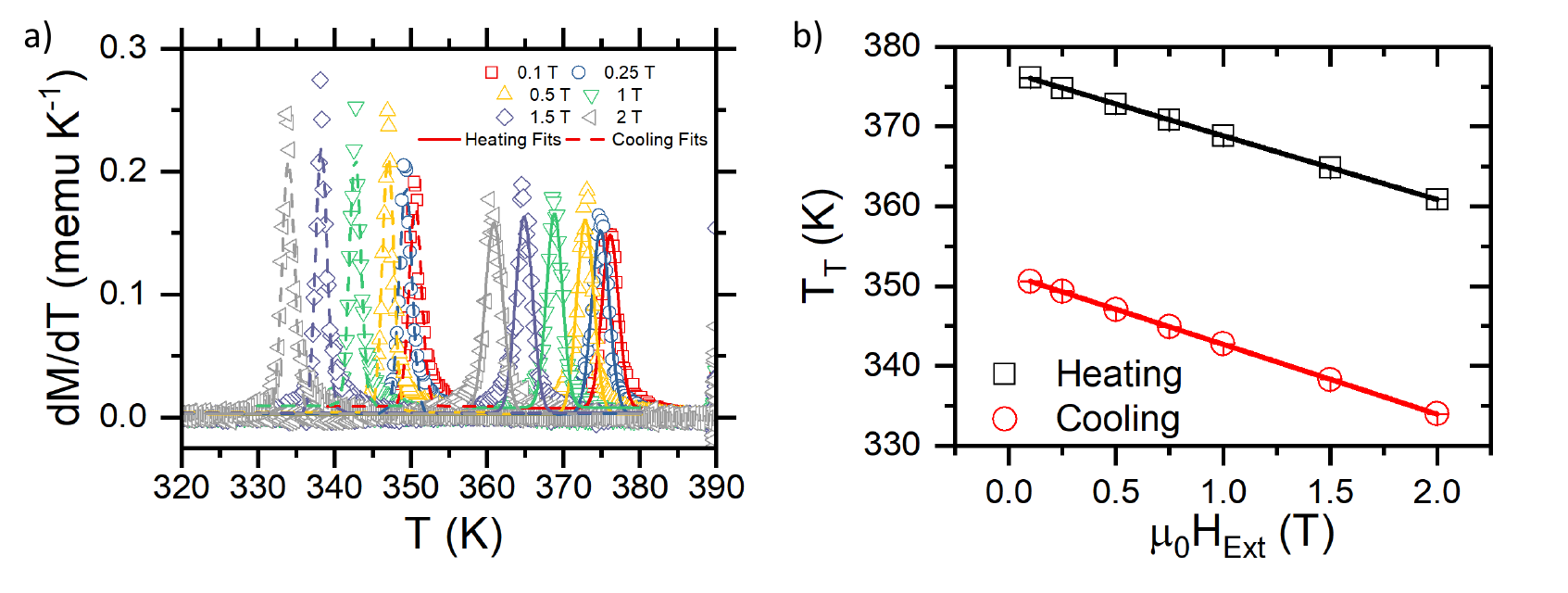}
    \caption{Dependence of the transition temperature, $T_\text{T}$ on external magnetic field. Panel (a) shows the dependence of the temperature, $T$, derivative of the magnetization, $dM/dT$, against $T$ for measurements performed in various external magnetic field strengths (points). The solid and dashed lines in this figure are Gaussian fits to the data performed when heating and cooling respectively. Panel (b) shows the value of the extracted values of $T_\text{T}$ against the external magnetic field in which the measurements was performed. The linear fit here is used to extract $dT_\text{T}/d(\mu_0 H_\text{Ext})$ and the transition midpoint at zero field, $T_\text{M}$. }
\label{fig:Supp_TT}
\end{figure*}

The dependence of the transition upon varying magnetic field values can be seen by the points in Fig.~\ref{fig:Supp_TT}(a). This figure shows the temperature derivative of the magnetization $dM/dT$ for a series of field values, where the position of the peak in $dM/dT$ clearly decreases in temperature with increasing field. The value of $T_\text{T}$ is extracted by fitting a Gaussian curve to $dM/dT$ \cite{Maat2005}, which can be seen by the dashed and solid lines in this figure for measurements performed when cooling and heating, respectively. The extracted values of $T_\text{T}$ are shown against their respective value of $\mu_0 H_\text{Ext}$ for both the heating and cooling branches in Fig.~\ref{fig:Supp_TT}. The linear fit here is used to extract $dT_\text{T}/d(\mu_0 H_\text{Ext})$ and also the transition midpoint at zero field, $T_\text{M}$, as the absiccsa intercept. Values extracted for these measurements are shown in Table \ref{table:Supp_TT}.

\begin{table}[b]
	\caption{Results of linear fits to the dependence on $T_\text{T}$ on $\mu_0 H_\text{Ext}$.
		\label{table:Supp_TT}}
		\begin{center}
        \begin{tabular}{ |c|c|c| }
        \hline
        & $\frac{dT_\text{T}}{d(\mu_0 H_\text{Ext})}$ (K T$^{-1}$) & $T_\text{M}$ (K) \\
        \hline
        Heating & $-(7.99 \pm 0.02)$ & $376.85 \pm 0.02$ \\
        Cooling & $-(8.76 \pm 0.03)$ & $351.47 \pm 0.03$ \\
        \hline
        \end{tabular}
        \end{center}
\end{table}

\section{X-Ray Absorption Spectroscopy Results}
\begin{figure}[t]
    \includegraphics[width = 16cm]{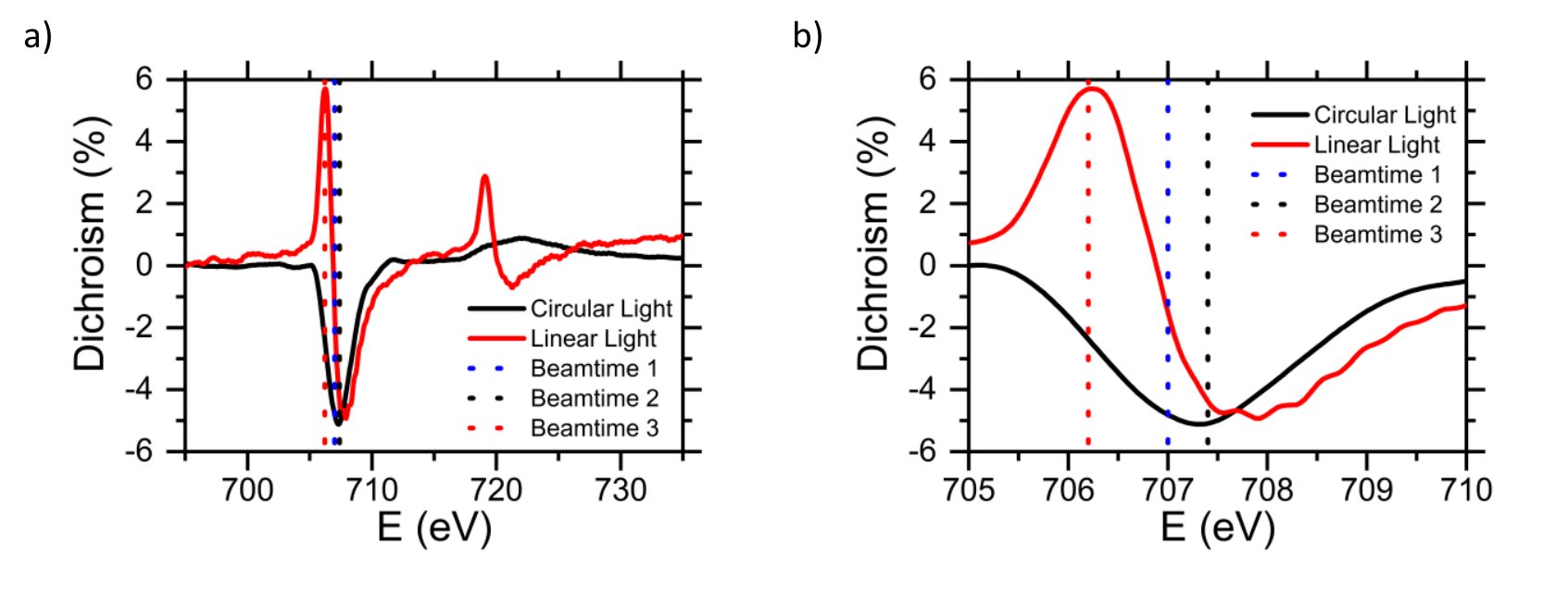}
    \caption{X-ray absorption spectroscopy results. (a) The value of $D$ against beam energy for both linearly (solid black line) and circularly (solid red line) polarized light taken at 400 K. The dashed lines here show the position of the beam energy used in each of the three beamtimes that make up the XPCS results. (b) a close up of the scan shown in (a) between 705 and 710 eV. Beamtimes 1 and 2 focussed on measurements performed using XMCD so should be judged against the black solid line. Whilst the XMLD measurements were all performed in the 3rd beamtime and so should be compared to the red solid line. \label{fig:XAS}}
\end{figure}

The location of the Fe L$_3$ resonance edge was identified using X-Ray absorption spectroscopy. In this, an energy scan from 690 to 730 eV was performed using both helicities of circular light and both polarizations of linear light $(I^+, I^-)$ and calculating the dichroism, $D$, using,
\begin{equation}
D = \frac{I^+ - I^-}{I^+ + I^-}. 
\end{equation}
The results for both circular (black solid line) and linear (red solid line) light for scans performed at 400 K are shown in Fig.~\ref{fig:XAS}. The presence of dichroism is clearly evident around 707 eV for both dichroism types here.

Also, shown on this figure is the positions of the beam energy (dashed lines) that was used in each of the three beamtimes that made up the data presented in the main paper. To make this easier to see, a close up of the energy scan between 705 and 710 eV is shown in Fig.~\ref{fig:XAS}(b). Beamtimes 1 and 2 focussed on measurements performed using XMCD so should be judged against the black solid line. Whilst the XMLD measurements were all performed in the 3rd beamtime and so should be compared to the red solid line. 

\section{Demonstration of Resonant Magnetic Small Angle Scattering}
\begin{figure}[t]
    \includegraphics[width = 8cm]{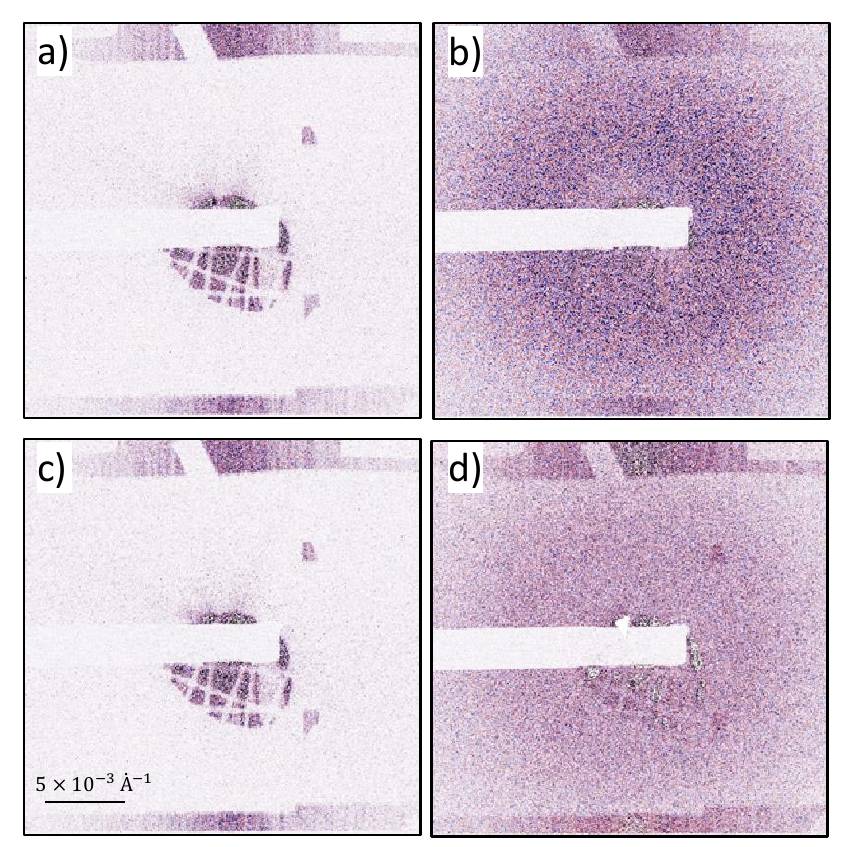}
    \caption{Transmission RMSAXS patterns. (a) Image taken with a beam energy of 690~eV--off-resonance--at 400 K using circularly polarized light, showing no appreciable scattering. (b) Image taken on the Fe L$_3$ resonance edge using circularly polarized light, showing a clear SAXS ring with speckle. Equivalent images for linearly polarized light are shown in panels (c) and (d). In all four images the shadow cast by the beamstop is visible as a white rectangle on the left, and the pinhole and crossed TEM grids are visible in the centre of each image. \label{fig:x2}}
\end{figure}

Evidence for the presence of resonant magnetic small angle scattering can be seen in panels (a)-(d) of Fig.~\ref{fig:x2}. These images are examples of those taken with the beam energy both off (panels (a) and (c)) and on (panels (b) and (d)) the Fe L$_3$ resonance for images taken using circularly (panels (a) and (b)) and linearly (panels (c) and (d)) polarized light. These images are taken using the protocol outlined in the methods section of the main document. Both images taken with beam energies away from the Fe L$_3$ edge show no appreciable scattering of any kind, whilst those taken on the Fe L$_3$ edge clearly show a small angle scattering ring with speckle features. The resonant enhancement of the scattering proves that it originates from magnetic order within the sample. The presence of speckle indicates the presence of a disordered magnetic structure, which is consistent with the expected magnetic domain structure at this point in the transition \cite{Almeida2017, Temple2018}.

\section{Behaviour of the Speckle Amplitude and Mixing Frequency}
\begin{figure}[t]
    \includegraphics[width = 13cm]{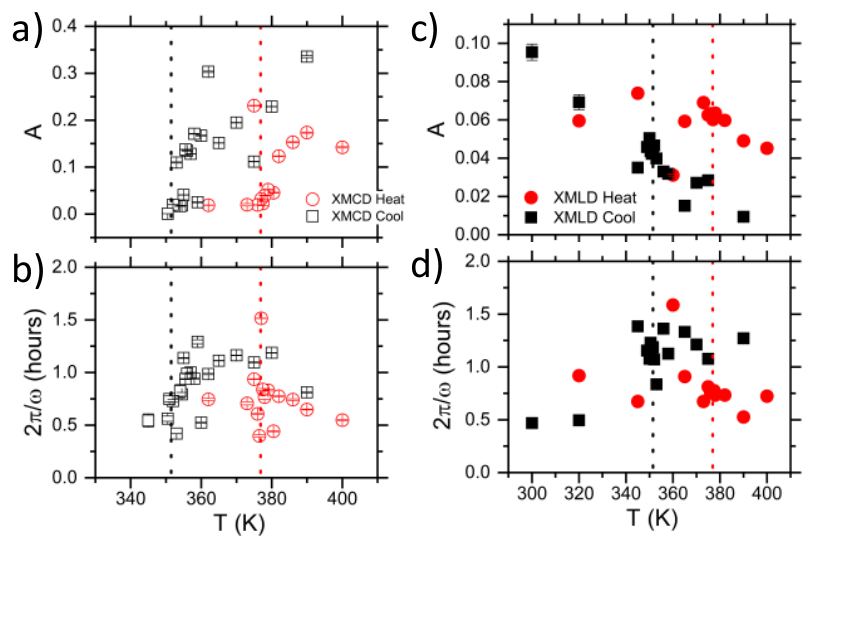}
    \caption{Temperature dependence of the speckle amplitude, $A$, and time constant associated with the mixing frequency, $2\pi/\omega$ for all measurements. (a) and (c) show the temperature dependence of $A$ for all measurements  extracted for the fits of the stretched exponent model to the $g_2$ curves for the XMCD and XMLD measurements, respectively. (b) and (d) show the extracted values of $2\pi/\omega$ for the XMCD and XMLD measurements. The dashed lines show the position of the transition midpoint measured using magnetometry. }
\label{fig:Supp_Aw}
\end{figure}

The behaviour of the speckle amplitude, $A$, and the time constant associated with the mixing frequency $2\pi/\omega$, extracted from the fits of the stretched exponential model to the $g_2$ curves are shown against temperature for both the XMCD and XMLD measurements are shown in Fig.~\ref{fig:Supp_Aw}. 

For the XMCD measurements, the extracted value of $A$, shown in Fig.~\ref{fig:Supp_Aw}(a), decreases with temperature from $\sim 0.2 \to \sim 0$ when cooling and increases between the same values when heating. This behaviour consistent with the changes in the volume of FM domains through the transition. The values of $A$ extracted from the XMLD measurements, shown in Fig.~\ref{fig:Supp_Aw}(c), shows a quasi-linear increase between $A\sim 0.01 - 0.1$ with decreasing temperature when cooling. The behaviour when heating for the XMLD measurements is consistent at around $A \sim 0.06$ for measurements performed below the transition midpoint (dashed lines). For measurements performed for $T > T_\text{M}$ when heating, $A$ decreases with temperature from $\sim 0.06$, reaching $\sim 0.05$ at 400 K. 

The behaviour of the time constant associated with the mixing frequency, $2\pi/\omega$, extracted from the XMCD measurements is shown in Fig.~\ref{fig:Supp_Aw}(b). When cooling, $2\pi/\omega$, decreases between 1.25 and 0.5 hours. When heating, for measurements performed below $T_\text{M}$, there appears to be a large scatter in the data which extracted values ranging between 1.5 and 0.25 hours. For measurements where $T> T_\text{M}$, the value of $2\pi/\omega$ extracted decreases from $\sim 1$ to $\sim 0.5$ hours. The variations in $2\pi/\omega$ do not follow the behaviour of $\lambda$ shown in the main paper, which suggests that both the charge and magnetic components of the scattering from the FM regions exhibit fluctuations. The reason for this unknown at this stage.

The time constant extracted for the XMLD measurements appears to be largely invariant at around $\sim 0.75$ hours for the measurements ranges here with no clear trend for the measurements performed whilst heating. The XMLD measurements taken when cooling however, appear to dip at the transition midpoint from $\sim 1.5$ hours to $\sim 0.8$ hours. The variations in the data here appear to follow the behaviour of $\lambda$ shown in the main paper, suggesting that the charge fluctuations probed with XMLD are constant. 

\section{Dependence of the Dynamic Behaviour on the Extracted Lengthscale}
\begin{figure}[t]
    \includegraphics[width = 13cm]{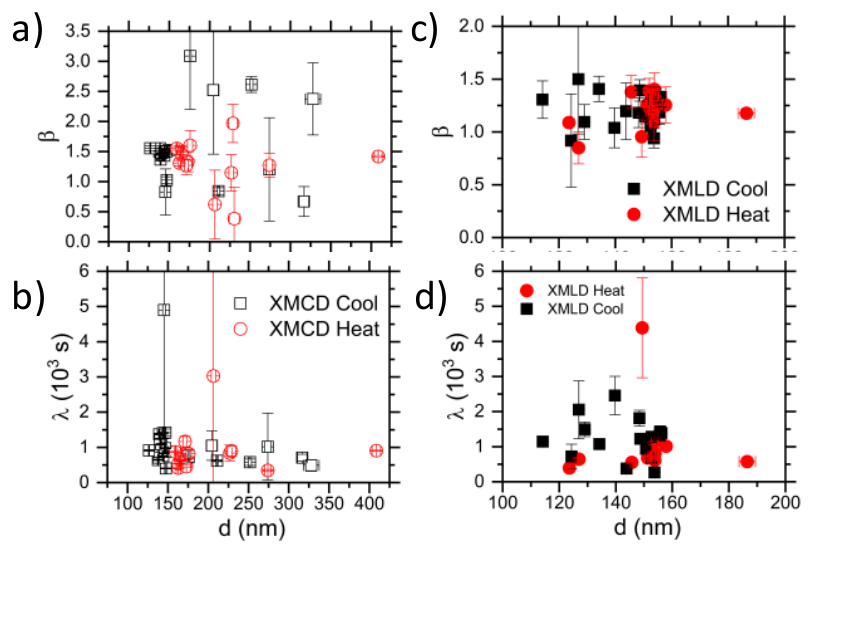}
    \caption{Dependence of the dynamic behaviour on extracted lengthscale. Panels (a) and (b) show dependence of the values of the stretching exponent, $\beta$, and the relaxation time, $\lambda$, extracted from the analysis of the temporal correlation functions against the extracted length scale, $d$, for the measurements performed using XMCD. Panels (c) and (d) shows the same analysis for the measurements performed using XMLD. There is no clear correlation between either $\beta$ or $\lambda$ on $d$ for any of the measurement sets here.}
\label{fig:Supp_d}
\end{figure}

As explained in the main body of this work, the dynamic behaviour of the system can sometimes be reliant upon the length scales present in the system \cite{ChenJam2013, Djurberg1997, Morley2017}, particularly in the case of critical scaling. The values of the stretching exponent, $\beta$, and the relaxation time, $\lambda$, extracted from the analysis of the temporal correlation functions are shown in Fig.~\ref{fig:Supp_d} against the extracted length scale, $d$, for all measurements. Measurements performed using XMCD are shown in panels (a) and (b), whilst measurements performed using XMLD are shown in panels (c) and (d). It is clear from this figure that there is no real correlation or dependence of either of the quantities governing the dynamic behaviour upon $d$. Therefore, we are unable to reconcile the changes in the dynamic behaviour seen in this work with the changes in the nature of the scatterer through the transition.

\section{Magnetic Viscosity Measured Using a Magnetometer Under Varying Conditions}
\begin{figure}[t]
    \includegraphics[width = 16cm]{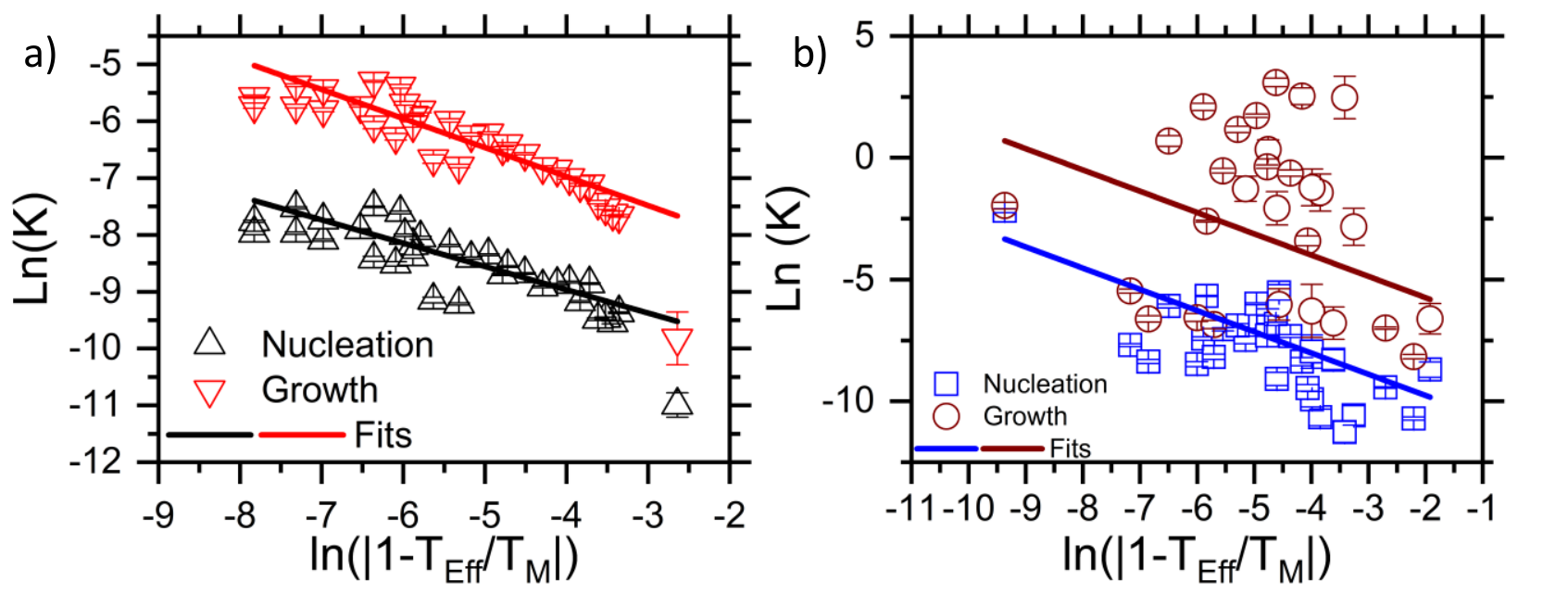}
    \caption{Results of Avrami analysis for magnetometer based magnetic viscosity measurements performed under varying conditions. Panel (a) shows the values of $\ln (K)$ extracted for fits to measurements performed when heating, for both the nucleation and growth regimes, plotted against $\ln |1-T_\text{Eff} / T_\text{M}|$. Panel (b) shows the same quantities extracted from the fits to measurements performed when cooling. The solid lines in these figures show linear fits in conjunction with the critical slowing down model presented in the main text. All data sets here present evidence of critical speeding up.}
\label{fig:Supp_OR}
\end{figure}

The results of the magnetic viscosity measurements performed using a magnetometer presented in the main body of the text are those that best replicate the conditions used in the XPCS experiment with regards to the temperature sweep rate. In these measurements a field is required to prevent artefacts being introduced into the data. Also performed were measurements of the same sample that use a temperature sweep rate of $10$ K min$^{-1}$ with no external magnetic field applied. These are found to have two regimes of development as consistent with those presented in the main body of the text. After being analysed in the same way as those in the main document, the extracted values of $\ln (K)$ are plotted against $\ln |1-T_\text{Eff} / T_\text{M}|$ in Fig.~\ref{fig:Supp_OR} for measurements performed whilst both heating (panel (a)) and cooling (panel (b)). In conjunction with the critical slowing down model presented in the main text, a linear fit is applied to extract the critical scaling exponent, $zv$, for each data set, the results of which are presented in Table \ref{table:Supp_OR}. Critical speeding up is observed for all data sets here.

\begin{table}[b]
	\caption{Results of critical scaling exponent, $zv$, extracted by fitting of the critical slowing down model to the magnetometer based magnetic viscosity measurements taken in the absence of magnetic field.
		\label{table:Supp_OR}}
		\begin{center}
        \begin{tabular}{ |c|c|c| }
        \hline
        & Nucleation & Growth \\
        \hline
        Heating & $-(0.41 \pm 0.04)$ & $-(0.51 \pm 0.04)$ \\
        Cooling & $-(0.9 \pm 0.1)$ & $-(0.9 \pm 0.3)$ \\
        \hline
        \end{tabular}
        \end{center}
\end{table}

\section{Behaviour of the Avrami Exponent}
\begin{figure}[t]
    \includegraphics[width = 16cm]{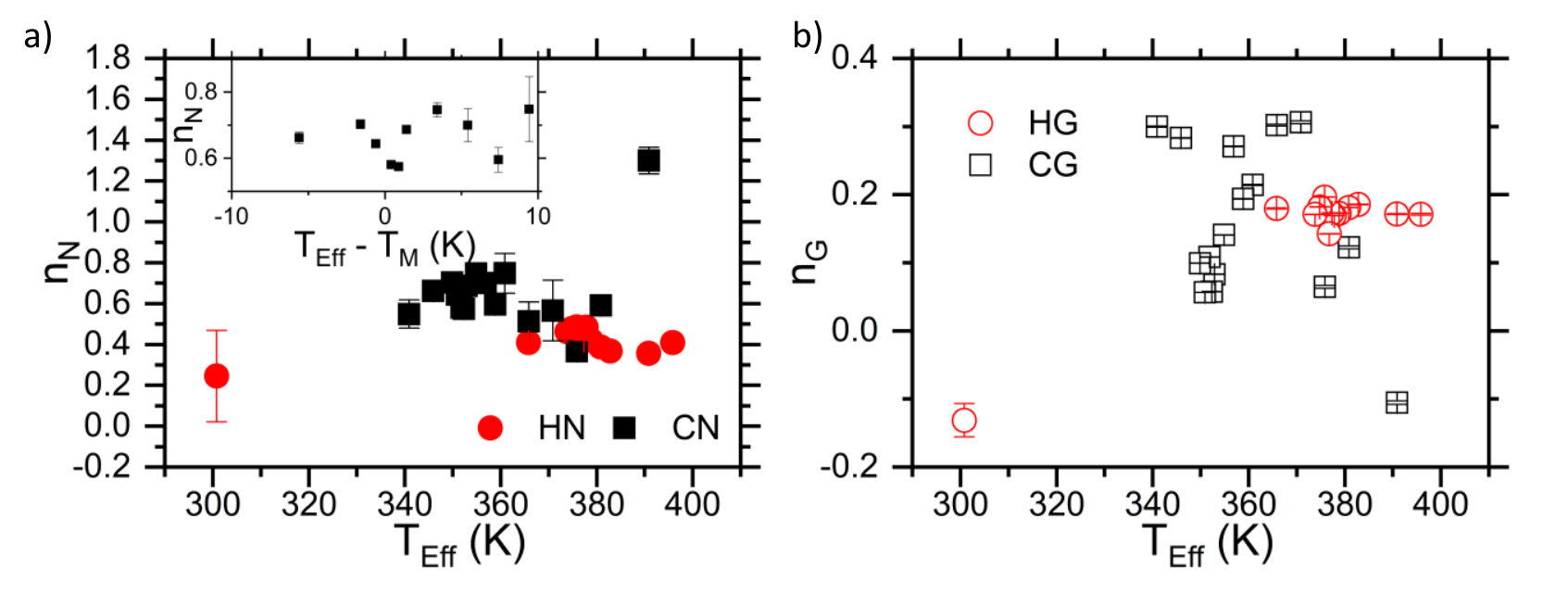}
    \caption{Behaviour of the Avrami exponent through the transition. Panel (a) shows the behaviour of the Avrami exponent, $n$, extracted for the measurements performed using a magnetometer presented in the main body of the work for the nucleation (N) regime. The inset in panel (a) shows the behaviour of $n$ centered around the transition midpoint, $T_\text{M}$. Panel (b) shows the same quantities but extracted from the fits to the growth regime. Red points show measurements performed when heating whilst black points are used to show measurements when cooling. All measurement sets performed when cooling see a variation in the extracted value of $n$ for $T_\text{Eff} \sim T_\text{M}$. There is no obvious change around this temperature for measurements performed when heating.}
\label{fig:Supp_n}
\end{figure}

The Avrami analysis shown in the main body of the text contains a constant, $n$, which is known as the Avrami constant \cite{LovingThesis}. The extracted value of $n$ for both the nucleation (N) and growth (G) phases of the dynamic behaviour is shown for all measurements presented in the main body of the text in Fig.~\ref{fig:Supp_n}(a) and (b) respectively. The extracted values of $n$ are typically between 0 and 1 and appear to vary again around the transition midpoint, for all measurement sets apart from the growth phase when heating. The cooling branch measurements see the value of $n$ decrease for $T \sim T_\text{M}$ for both types of phase kinetics here, whilst the heating branch measurements see no obvious change over the measurement range for both phases of the dynamics. 

The Avrami exponent is linked to the number of dimensions in which the changes within the sample take place \cite{LovingThesis,Lin2014}. Avrami exponents of less than 1 have been attributed to the influence of more than characteristic relaxation time for a given kinetic process \cite{Lin2014}. The change in the value of $n$ approaching $T_\text{M}$ then implies that the number of relaxation times changes at this temperature. This is likely to be due to the reduction in the relative energy barrier approaching this temperature and the distribution of transition temperatures through the film giving rise to a number of different environments each with their own kinetic properties. 

\section{Dynamic Investigations Through the Second-Order Ferromagnet to Paramagnet Phase Transition at the Curie Temperature}
\begin{figure}[t]
    \includegraphics[width = 16cm]{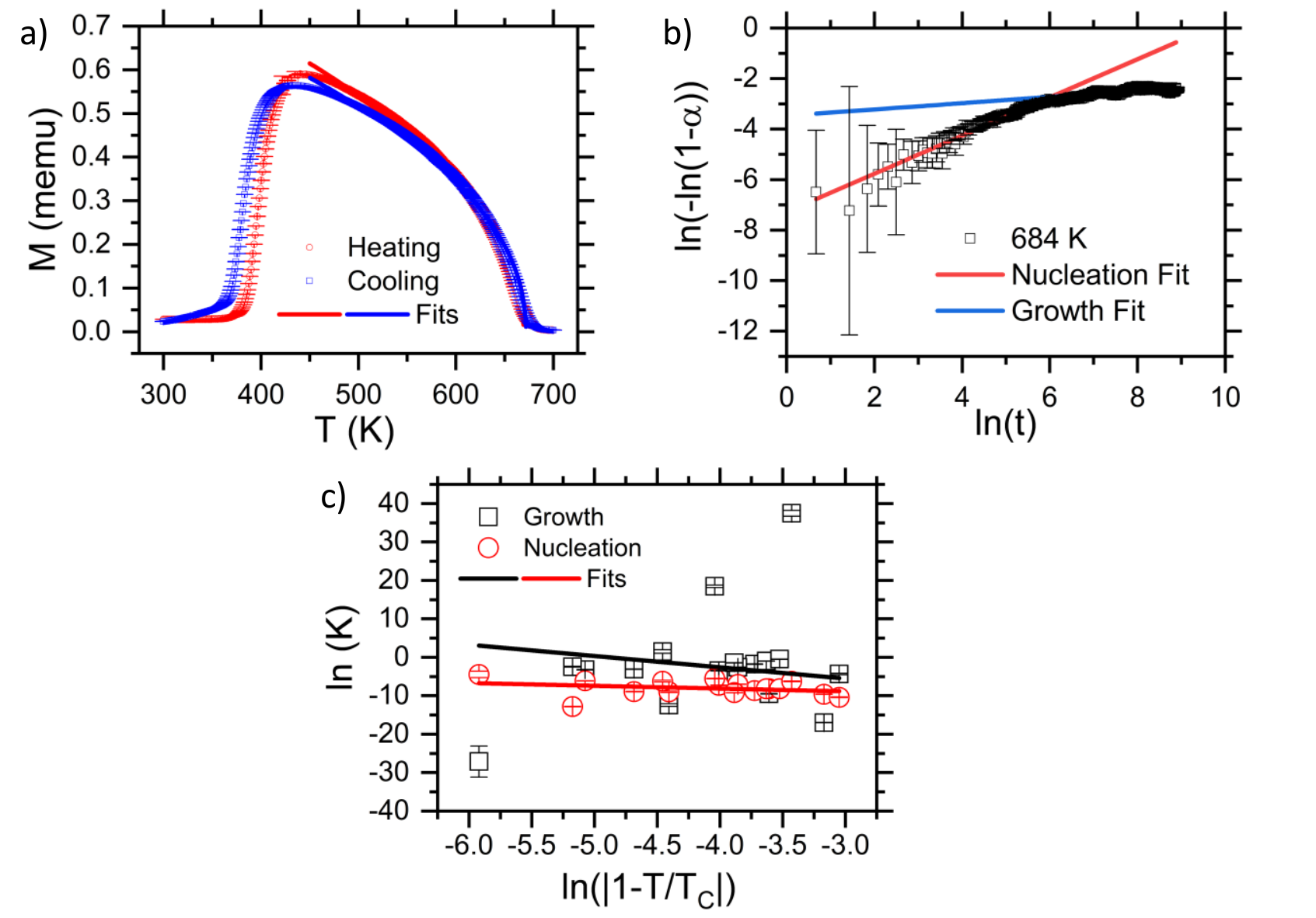}
    \caption{Magnetometry based magnetic viscosity measurements through the second-order phase transition at the Curie temperature, $T_\text{C}$. Panel (a) shows the magnetization against temperature profile of a 60 nm FeRh sample grown on MgO between 300 - 700 K performed in a 0.1 T externally applied magnetic field. The blue points are those taken whilst cooling, and the red points are those taken when heating. Panel (b) shows examples of the Avrami analysis for a measurements performed at 684 K after cooling from 720 K. The solid lines show the fits for the two distinct regions of dynamic behaviour. Panel (c) shows the extracted values of $\ln (K)$ from the Avrami analysis plotted against $\ln |1-T / T_\text{C}|$ for measurements performed when cooling through the second-order phase transition at $T_\text{C}$. A weak dependence on the proximity to the phase transition is seen here. }
\label{fig:Supp_SOPT}
\end{figure}

Further to the investigations focused on the behaviour of the system in proximity to the first-order transition (FOPT) temperature presented in the main body of this work, similar investigations were also performed through the second-order phase transition (SOPT) taking place at the Curie temperature, $T_\text{C}$. These were performed in the magnetometer in an externally applied field of 0.1 T. Due to the fragility of the sample used in the XPCS experiments these investigations are performed on a different 60 nm thick FeRh sample which is grown on MgO. The temperature dependent magnetization profile of the sample taken between 300-700 K is shown in Fig.~\ref{fig:Supp_SOPT}(a) and shows a FOPT between 300-450 K, as well as a SOPT around 670 K. To identify the exact position of $T_\text{C}$, the data between 450 and 700 K is fitted to the following equation \cite{mftbeta},
\begin{equation}
    M = M_0 \bigg(1 - \frac{T}{T_\text{C}}\bigg)^\beta,
\end{equation}
where $M_0$ is the magnetization at 0 K and $\beta$ denotes the nature of the approach towards the SOPT. The fits here yield values of $\beta = 0.493 \pm 0.001$ and $\beta = 0.467 \pm 0.001$ for fits to the heating and cooling branches respectively, which gives reasonable agreement with the mean-field model as consistent with previous measurements \cite{Massey2019}. The values of $T_\text{C} = 669.6 \pm 0.1 K$ and $T_\text{C} = 671.8 \pm 0.1$ K were extracted for the heating and cooling branches respectively.

Again, performing the Avrami analysis on this system reveals two distinct regions where a linear relationship between $\ln(-\ln(1-\alpha))$ is observed as seen for the measurements performed at 684 K in Fig.~\ref{fig:Supp_SOPT}(b), which is again attributed to the nucleation and growth of domains. The fits to the data for the two regions are shown by the solid lines in this figure. Fig.~\ref{fig:Supp_SOPT}(c) shows the behaviour of $\ln(K)$ extracted from the Avrami analysis for the measurements performed when cooling through the SOPT against $\ln |1-T/ T_\text{C}|$. Here, as all of the measurements are performed in the same applied field there is no need to correct the temperature. A weak dependence on the rate constant upon the proximity to $T_\text{C}$ is seen here, with the values extracted when both heating and cooling towards the SOPT shown in the summary of extracted values of $zv$ in the main text. Only one of the measurements (growth phase for heating) demonstrates a value of $zv$ that is significantly different from 0 within a 95\% confidence level.

\bibliographystyle{naturemag}

%\bibliography{Massey_thesis}